\documentclass[pre,aps,twocolumn,superscriptaddress,nopacs,floatfix,amssymb,amsmath]{revtex4}
\usepackage{epsfig}
\usepackage{graphics}
\newcommand{\beqn}{\begin{eqnarray}}
\newcommand{\eeqn}{\end{eqnarray}}
\newcommand{\eq}[1]{(\ref{#1})}

\begin{document}

\title{Elastic Energy and Phase Structure in a Continuous Spin Ising Chain \\
with Applications to the Protein Folding Problem}

%\vskip 5.0cm
\author{M.N. Chernodub}
\email{  Maxim.Chernodub@lmpt.univ-tours.fr }
\thanks{\\ On leave from ITEP, Moscow, Russia}
\affiliation{Laboratoire de Math\'ematiques et Physique Th\'eorique,
Universit\'e Fran\c{c}ois-Rabelais Tours, F\'ed\'eration Denis Poisson - CNRS,
Parc de Grandmont, 37200 Tours, France}
\affiliation{Department of Mathematical Physics and Astronomy,
Krijgslaan 281, 59, Gent, B-9000, Belgium}
\author{Martin Lundgren}
\email{Martin.Lundgren@physics.uu.se}
\affiliation{Department of Physics and Astronomy, Uppsala University,
P.O. Box 803, S-75108, Uppsala, Sweden}
\author{Antti J. Niemi}
\email{Antti.Niemi@physics.uu.se}
\affiliation{Department of Physics and Astronomy, Uppsala University,
P.O. Box 803, S-75108, Uppsala, Sweden}
\affiliation{Laboratoire de Math\'ematiques et Physique Th\'eorique,
Universit\'e Fran\c{c}ois-Rabelais Tours, F\'ed\'eration Denis Poisson - CNRS,
Parc de Grandmont, 37200 Tours, France}

\begin{abstract}
We present a numerical Monte Carlo analysis of a continuos spin Ising chain
that can  describe the statistical proterties of folded proteins. 
We find that depending on the value of the 
Metropolis temperature, the model displays the three known nontrivial phases  of
polymers:  At low temperatures the model is in a collapsed phase, at medium temperatures it is
in a random walk phase, and at high temperatures it enters the self-avoiding random walk
phase. By investigating the temperature dependence of the specific energy we confirm that
the transition between  the collapsed phase and the random walk phase is a phase transition, while
the random walk phase and self-avoiding random walk phase are separated from each
other by a cross-over transition. We also compare the predictions of the model to 
a phenomenological elastic energy formula, proposed by Huang and Lei to describe folded 
proteins. 
\end{abstract}

\pacs{???}

\date{22-07-2010}

\maketitle

\section{Introduction}

The concept of universality  \cite{wilson}, \cite{kadanoff} divides critical physical
systems  into universality classes that differ from each other essentially only 
 by their space-time dimensionality 
and the symmetry group of their order parameter. This enables the
computation of  critical properties  for an entire class of physical systems
using only a single representative model. In the case 
of polymers one expects that there are three different  {\it nontrivial} 
phases and these correspond to  the universality class of self-avoiding random
walk (SARW), to the universality class of Brownian motion  {\it i.e.} ordinary random walk (RW),
and to the universality class of  polymer collapse  \cite{degennes1}. 
These phases are each characterized  by the different
values of certain critical exponents  that describe the scaling properties of the 
polymer in the limit where the number $N$ of monomers becomes large. 
The most widely
used critical exponent, the compactness index $\nu$,  computes the inverse of the Hausdorff dimension
of the polymer. It can be introduced by considering how the polymer's
radius of  gyration $R_g$  increases in the number of monomers, asymptotically
for large values of $N$  \cite{macro},
\begin{equation}
R^2_g \ = \ \frac{1}{2N^2}  \sum_{i,j} ( {\bf r}_i 
- {\bf r}_j )^2 \  \approx \  R_0^2 N^{2\nu} ( 1 + \beta_1 N^{-\Delta_1} + ... ) 
\label{nu}
\end{equation} 
Here ${\bf r}_i$ ($i=1,2,...,N$) are the locations of the $N$
monomers in $\mathbb R^3$. The critical exponents $\nu$ and $\Delta_1$ are universal quantities.
But the form factor $R_0$ that characterizes the effective distance between the monomers in the
large $N$  limit, and  the amplitude $\beta_1$ that parametrizes  the leading finite size corrections, are not. 
The asymptotic  expansion (\ref{nu}) is an example of a general result \cite{wegner}, \cite{sokal} that states,
that when $N$ becomes large the mean value of any global observable $\mathcal O$ of a polymer 
should behave like
\[
<{\mathcal O}>_N \ = \ A N^p \left[ 1 + \frac{\alpha_1}{N} + \frac{\alpha_2}{N^2} + \dots \right.
\]
\begin{equation}
\left. + \frac{\beta_1}{N^{\Delta_1}} + \frac{\beta_2}{N^{\Delta_1 + 1}}
+ \frac{\beta_3}{N^{\Delta_1 + 2}} + \dots \right]
\label{genobs}
\end{equation}
where the exponents are universal, but the pre-factor and the various amplitudes are all non-universal.

For a polymer the  compactness index has the following mean field (mf)
values \cite{degennes1}: 
\begin{equation}
\nu_{\rm mf} \ = \ \left\{ \!\! \!\! \begin{matrix} 3/5 \ \   {\rm SARW} \\ 
\! \! \! 1/2 \ \ \  \ {\rm RW}
 \\
 \ \ \   \ 1/3 \ \  {\rm collapsed}
 \end{matrix} \right.
 \label{nus}
 \end{equation}
As a function of temperature, the collapsed phase occurs at low temperatures (bad solvent) while the
SARW describes the high temperature (good solvent) behavior of polymers. 
The random walk phase takes place at the $\Theta$-temperature  that 
separates the  SARW phase from the collapsed phase.  
In general the mean field values of the critical exponents 
acquire corrections due to fluctuations, and for the universality class of the self-avoiding random walk
the improved values are $\nu = 0.5880 \pm 0.0015 $ and $\Delta_1 = 0.47 \pm 0.03$. These values
were obtained in  \cite{Zinn} by utilizing the concept of universality that relates the self-avoiding random walk
with the $n\to 0$ component $\phi^4$ field theory \cite{degennes2}. The subsequent direct Monte Carlo 
evaluation reported in \cite{sokal} gave the very similar values $\nu = 0.5877 \pm 0.0006$ 
and $\Delta_1 = 0.56 \pm 0.03$, in line with the concept of universality.

Qualitatively, at the level of a mean field theory  the  phase structure of a polymer can 
be described in terms of the Flory-Huggins theory \cite{degennes1}. For this we characterize 
the polymer concentration by an order parameter $\phi(x)$, with  $0\leq \phi(x) \leq 1$.  At low concentrations
the polymer free energy 
density (per temperature) has the Landau expansion
\begin{equation}
\frac{1}{T} E [\phi]   =  \eta (\nabla \phi)^2 + \gamma \cdot \phi \ln \phi + \frac{1}{2} (1-2\chi) \phi^2 + \frac{g}{3!} \phi^3 + ... 
\label{FH}
\end{equation}
Here $\eta,  \gamma, \chi$ and $g$ are parameters. The first term is a stiffness term. The second term describes
entropy contrubutions.  The third term describes monomer-monomer interactions; the (Flory) interaction parameter
$\chi$ is generically a decreasing function of temperature. The last term characterizes the three-body (and
higher order contributions) monomer interactions.  The phase structure can be exposed by 
 by ignoring the stiffness term and by minimizing the remaining potential energy contribution to 
free energy. With proper  relative values of the parameters the potential has a form that is familiar
from spontaneous symmetry breaking: When the ground state expectation value  $<\!\! \phi\!\! >$ is non-vanishing we
are in the  collapsed phase while the vanishing value $<\!\! \phi \!\! > \approx 0$ implies that we are in  the universality
class of self-avoiding random
walk. The border line that separates these two phases determines the $\Theta$ temperature  where the polymer is in the
universality class of random walk. It  occurs at that value of temperature (or denaturant concentration) 
for which the excluded volume parameter vanishes,  and to first order
\[
1-2\chi(T_\Theta)  \ = \ 0
\]
Thus, for $\chi(T) > 1/2$ we are in the collapsed phase while for $\chi (T) < 1/2$ we enter the SARW phase and
in particular at the $\Theta$-point the $\phi^2$ ({\it i.e.} mass) contribution to the free energy is absent.

Here we shall present results of an extensive numerical analysis of the polymer phase structure.
Our approach is based on the chiral  homopolymer model introduced in \cite{ulf}. The 
applicability of the model to analyze the properties of 
chiral polymers in all three phases can be justified by the concept of universality.  Indeed, the derivation 
of the model in \cite{ulf} is very much based on the universality concept: The model 
accounts for the monomer complexity, the presence of amino acid side chains in proteins, 
 and polymer-solvant interactions in an effective manner. In particular,
the model appears to describe certain universal properties of the folded
proteins \cite{dill} in the Protein Data Bank (PDB) \cite{pdb} with a very high accuracy.
More recently, it has also been shown \cite{maxim}, \cite{nora} that the model  supports dark solitons and
the presence of these solitons appears to be related to the emergence of the collapsed phase.
These solitons can also describe folded proteins in PDB with a subatomic accuracy of less than 1 \.Angstr\"om in root 
mean square distance (RMSD).  This also motivates us to  compare our results  
to a recently presented phenomenological model of protein folding \cite{huang1}.

\section{The Model}
 
The model  introduced in \cite{ulf}  is defined by the following internal energy,
\beqn
E & = & \phantom{+} \underset{ij}{\sum} a_{ij}\left\{1-\cos\left[\omega_{ij}\left(\kappa_{i}-\kappa_{j}\right)\right]\right\}
\label{eq:model}\\
& &  + \underset{i}{\sum}\left\{ b_{i}\kappa_{i}^{2}\tau_{i}^{2}+c_{i}\left(\kappa_{i}^{2}-m_{i}^{2}\right)^{2}+d_{i}\tau_{i}\right\}
\nonumber
\eeqn
Here $i,j = 1, ..., N$ label  the $N$ monomers of a (chiral) polygonal chain in  $\mathbb R^3$. These monomers
are located at the vertices of the polygon, and the chain geometry 
changes when the polymer fluctuates in $\mathbb R^3$.  The geometry is determined by
the  order parameter $\kappa_i$ that is a discrete lattice version of the 
 Frenet curvature, and  by the order parameter $\tau_i$ that is
the lattice version of the Frenet torsion \cite{ulf}. Once the values of $(\kappa_i, \tau_i)$ for each $i=1,...,N$ 
are given the actual shape of the polymer as a polygonal chain in the three dimensional 
space can be computed by integrating the appropriate discrete version of the Frenet equations. This integration  
introduces parameters $\delta_i$, the three dimensional distances between the monomers.

The  $a_{ij}, \omega_{ij}, b_i, ... , d_{i}$ in (\ref{eq:model}) are parameters.
The first sum in the free energy describes long-distance interactions,
we have introduced the cosine function to tame excessive fluctuations 
in $\kappa_i$ in the numerical simulations. In the second sum the first term  describes the interaction between
$\kappa_i$ and $\tau_i$,  and the second term describes the self-interaction 
of $\kappa_i$. Finally, the last term is a discretized  version of the one dimensional Chern-Simons 
functional,  it  is the origin of  chirality in the polymer chain \cite{ulf}, with handedness that depends on the sign
of $d_i$.

For a general polymer the quantities ($a_{ij},\omega_{ij}, b_i, c_i,$ $\mu_i,d_i$) are {\it a priori} site-dependent
parameters,  and different values of these parameters can be used to describe 
different kind of monomer (amino acid) structures.  For generic  $a_{ij}$  (\ref{eq:model})
is a spin-class model. Here we shall be interested in the limiting case of a homopolymer
where  we restrict ourselves to only the nearest neighbor interactions
with
\begin{equation}
a_{ij} = \left\{  \begin{matrix}
\hskip -1.2cm 
a \cdot ( \delta_{i, i+1} + \delta_{i, i-1} ) \ \ \ \ (i=2, ... ,N-1)
\\
\hskip 0.2cm a \ \ \ \ \ \ \ \ (i=1,j=2) \ \ \& \ \ (i=N-1,j=N) \end{matrix}
\right.
\label{a}
\end{equation}
and we also select {\it all} the remaining parameters to be
{\it independent } of the site  index $i$.  Thus the model in the form studied here
reads
\[
E  =  \phantom{+} \underset{n.n.}{\sum} a \left\{1-\cos\left[\omega\left(\kappa_{i}-\kappa_{j}\right)\right]\right\}
\]
\begin{equation}
 + \underset{i}{\sum}\left\{ b \kappa_{i}^{2}\tau_{i}^{2}+c \left(\kappa_{i}^{2}-m^{2}\right)^{2}+d \tau_{i}\right\}
\label{eq:model2}
\end{equation}
where the first sum extends over the nearest neighbors; Notice that since the overall scale of the parameters $a,b,c$ and $d$ 
can be absorbed into the definition of the scale of the  Metropolis temperature $T$, as it stands there are  five independent intrinsic parameters.
Consequently the scale of energy, say in electronvolts,  remains indeterminate and should be defined by (re)normalization at some convenient value of $T$. 
We also note that classically, the model (\ref{eq:model2}) has a ground state which is a helix, with $\kappa_i \approx  \pm m$.

We select the numerical values
of the parameters in
a manner that allows for  a direct statistical comparison to PDB data. These values have been
found by a trial-and-error comparison with PDB data \cite{ulf}  and they are shown in
Table~\ref{tbl:parameters}. 
\begin{table}[!thb]
\begin{centering}
\begin{tabular}{|c|c|}
\hline
Parameter & Value\\
\hline
$a$ & 4\tabularnewline
$\omega$ & 4.25\tabularnewline
$b $ & $5.488 \cdot 10^{-4}$\tabularnewline
$c$ & 0.5\tabularnewline
$m$ & 24.7\tabularnewline
$d$ & -20\tabularnewline
\hline
\end{tabular}
\caption{Parameter values of the model~\eq{eq:model} that we use in our simulations}.
\label{tbl:parameters}
\par\end{centering}
\end{table}
Furthermore, we shall assume that the distances $\delta_i$ between the monomers that we need to introduce
when we integrate the discrete Frenet equations to construct the polygonal chain in $\mathbb R^3$, have the fixed value
\begin{equation}
|{\bf r}_ i - {\bf r}_{i-1} | = \delta \ = \  3.8 \  ({\buildrel _{\circ} \over {\mathrm{A}}} )   \ \ \ \ i=2,...,N .
\label{cond1}
\end{equation}
This value (in ${\buildrel _{\circ} \over {\mathrm{A}}}$) is chosen to coincide with the average distance
between $C_\alpha$ carbons in the backbone of PDB proteins. 
Finally, we exclude steric clashes by demanding that the distance 
between any two monomers  satisfies the bound
\begin{equation}
|{\bf r} _i - {\bf r}_j | \geq z  \ = \ 3.7 \ ({\buildrel _{\circ} \over {\mathrm{A}}}) \ \ \ {\rm   for} \ \ \ 
|i-j| \geq 2 .
\label{cond2}
\end{equation}
Again, this numerical value has been chosen to match the protein data in PDB.

We have used the standard Metropolis algorithm to simulate the model~\eq{eq:model}.
The initial configuration is a straight rod with $\kappa_i = \tau_i = 0$.
Each Monte-Carlo step consists of a shift of the curvature and torsion
by a typical value of $\Delta \kappa_i = \Delta \tau_i =0.05$. This shift  is accepted with
the probability 
\[
P=\min\left(1,\exp\left(-\frac{\triangle E}{T}\right)\right)
\]
where $T$ is the Metropolis temperature. We use this 
temperature as an external parameter 
that allows us to probe the different phases of the polymer.

The simulations proceeded as follows: For each temperature value, between 10
and 16 different polymer lengths was selected.
The number of the Monte-Carlo iterations of each chain was chosen to be
11.000 multiplied by the number $N$ of monomers in  the polymer. 
We created around  200 or more polymers for each individual temperature 
value $T$ and monomer number $N,$ with less for the extremely long and the 
highest temperature curves. The shortest polymers in our simulations had 50
monomers, and the longest ones had 1.800 monomers. These values were chosen
to be representative of the single domain proteins in PDB.

Finally, since the free energy (\ref{eq:model}), (\ref{eq:model2}) is quadratic in $\tau_i$ and furthermore
since $\tau_i$  only 
interacts locally, we can eliminate it by using its equation of motion
\begin{eqnarray}
\frac{\partial E}{\partial\tau_i} \ = \ 2b_i \kappa_i^2 \tau_i + d_i \ = \ 0 \nonumber \\ 
\Rightarrow \ \ \tau_i[\kappa_i] = - \frac{d_i}{2 b_i \kappa_i^2} \ \ \ 
\label{tk}
\end{eqnarray}
This gives us
\beqn
E & = &  \phantom{+}  \underset{i,j}{\sum} a_{ij} \left\{1-\cos\left[\omega_{ij}\left(\kappa_{i}-\kappa_{j}\right)\right]\right\}
\nonumber\\
& &  
+ \underset{i}{\sum}\left\{  c_{i}\left(\kappa_{i}^{2}-m_{i}^{2}\right)^{2}\ - \frac{d_i^2}{2b_i \kappa_i^2} \right\}
\nonumber
\eeqn
and in the limit of uniform chain and small $\omega$ we get  (after we add  boundary contributions and choose $\kappa_0 = \kappa_{N+1} = 0$)
\[
E \ \approx \ - a\omega^2 \sum\limits_{i}  \kappa_{i}\kappa_{i+1}  \ +
\]
\beqn
+ \sum\limits_i \left\{  a \omega^2 \kappa_i^2 +  c \left(\kappa_{i}^{2}-m^{2}\right)^{2}\ - \frac{d^2}{4b\kappa_i^2} \right\}
\label{eq:model3}
\eeqn
We recognize here 
a version of the continuos spin 
Ising chain \cite{ising}:  Indeed, the {\it only} difference between (\ref{eq:model3}) and the conventional continuous spin Ising chain
is in the presence of the last 
term in (\ref{eq:model3}). We note that this last term that has its origin in (\ref{tk}),  is quite reminiscent of the potential term 
that appears in the widely studied Calogero  model \cite{galo}, for the relative coordinate in the two-body case.
Furthermore, if we absorb the parameter combination 
$a\omega^2$ into the definition of overall scale of temperature , in (\ref{tk}), (\ref{eq:model3}) there are
only four  independent parameter combinations.

It  has been a commonly held  point
of view  \cite{barma}  that the lattice version of the $\phi^4$ model is always in the same universality class with the pure 
Ising model. But this has been disputed in the one dimensional case by explicit computations {\it e.g.} in \cite{baker}. 
Here we have an additional interaction term, the last Calogero-type term in (\ref{eq:model3}), and we shall explicitely 
show that the ensuing phase structure is highly nontrivial.

\section{The radius of gyration}

We shall first investigate the radius of gyration (\ref{nu}) with the goal to confirm that
the model \cite{ulf} does indeed describe the three different polymer phases characterized by the 
mean field values (\ref{nus}) of the critical exponent $\nu$. 

\begin{figure}[!thb]
\begin{center}
\begin{tabular}{c}
\includegraphics[scale=0.43,clip=true]{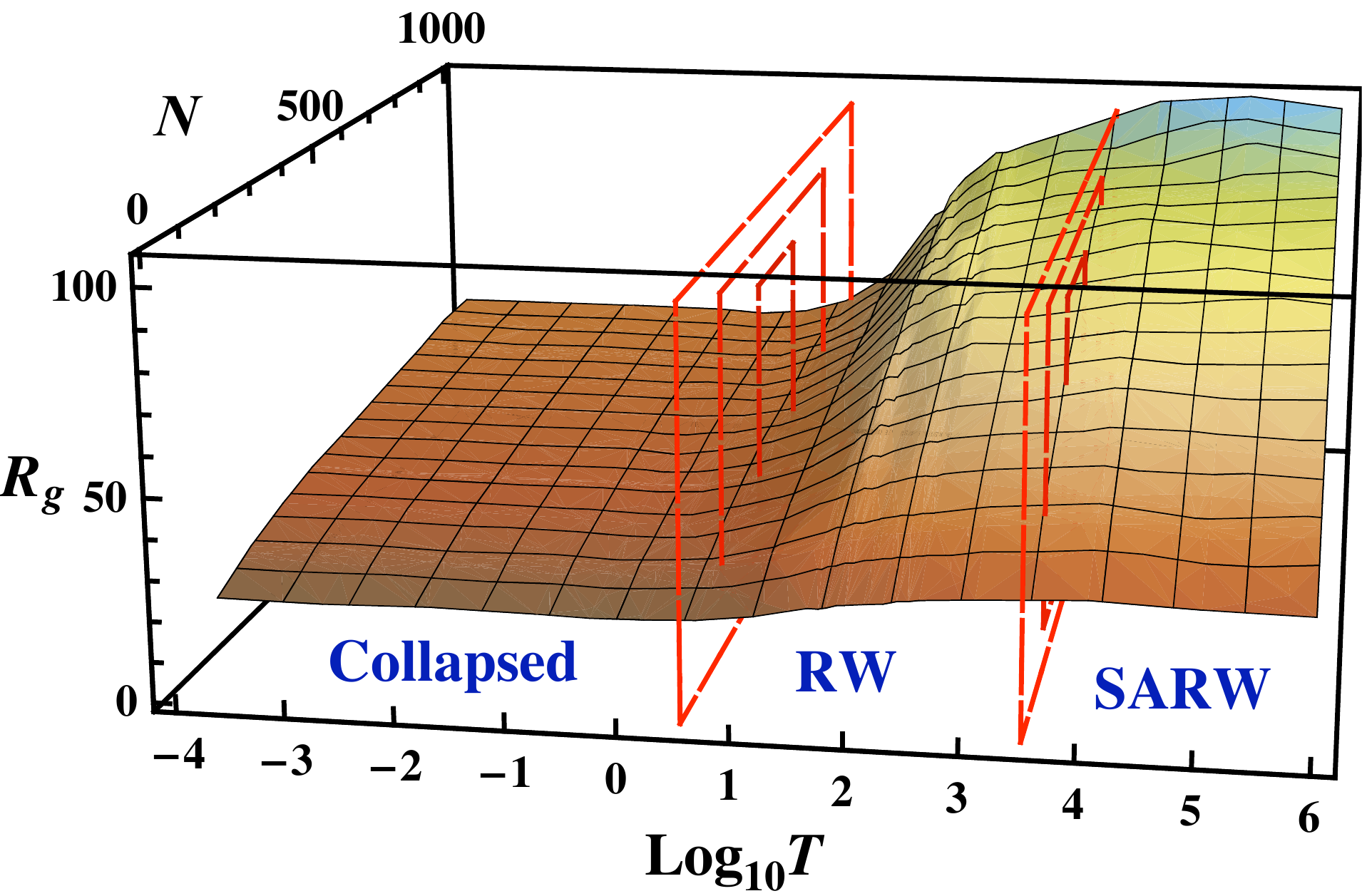}
\end{tabular}
\end{center}
\caption{The  radius of $R_g$ as a function of temperature $T$ and the number of monomers $N$. 
The three putative phases are identified with the putative 
position of the ensuing critical temperatures,  denoted by the vertical planes. 
}
\label{fig:Rg:3d}
\end{figure}
\begin{figure}[!thb]
\begin{center}
\includegraphics[scale=0.43,clip=true]{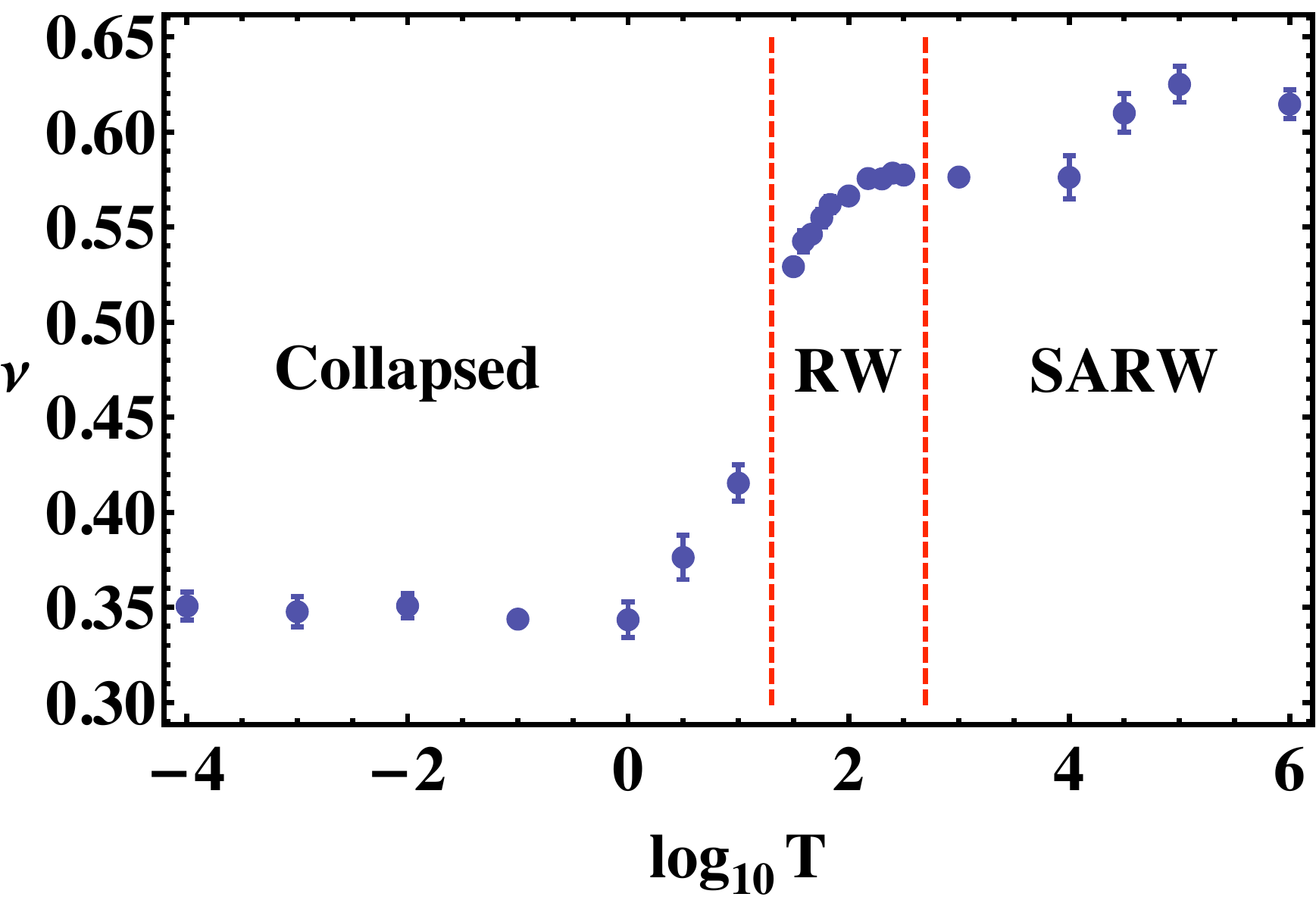}
\end{center}
\caption{The compactness index $\nu$  {\it vs.} temperature $T$. The vertical lines correspond to
temperature values where $\nu(T)$ reaches the mean field values (\ref{nus}).
}
\label{fig:nu}
\end{figure}
\begin{figure}[!thb]
\begin{center}
\includegraphics[scale=0.43,clip=true]{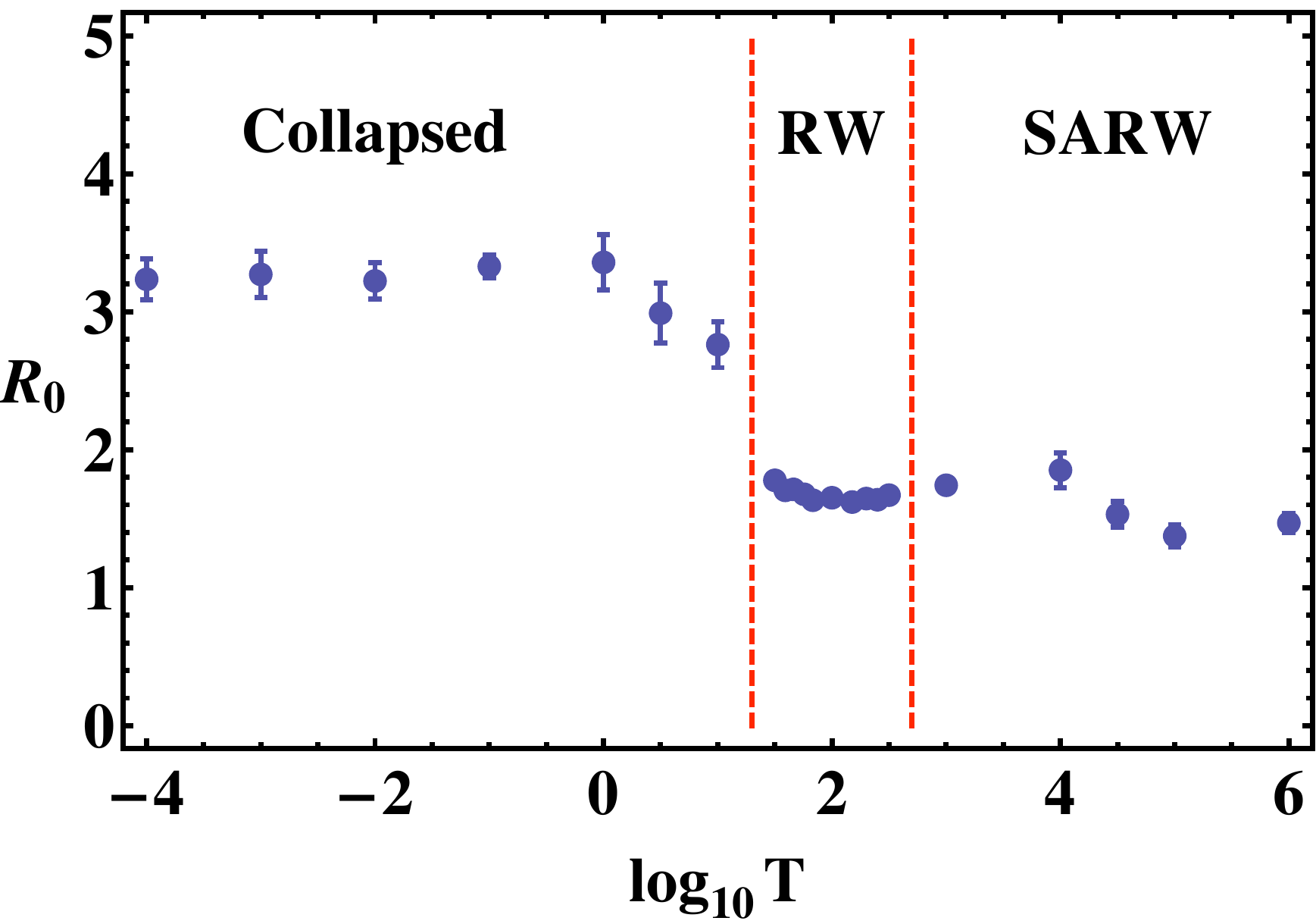}
\end{center}
\caption{The pre-factor $R_0$  in (\ref{nu}) {\it vs.} temperature $T$. The vertical lines correspond to
the temperature values where $\nu(T)$ in figure \ref{fig:nu} reaches the mean field values (\ref{nus}).}
\label{fig:R0}
\end{figure}

In Figure \ref{fig:Rg:3d} we show how the radius of gyration $R_g$ depends on the (Metropolis) temperature $T$
and the number of monomers $N$.  In Figure \ref{fig:nu} we depict the $T$ dependence of $\nu$,
and figure \ref{fig:R0} shows the $T$ dependence of  the pre-factor $R_0$ in (\ref{nu}).

\subsection{Collapsed phase}

In the figure \ref{fig:Rg:3d} we clearly identify a low-temperature phase which is the putative collapsed phase.   
In this phase $R_g$ is constant, or has only  very weak $T$ dependence, and 
we can fit the data with very high accuracy using
the following relation,
\begin{equation}
R_{g} = R_{0} N^{\nu}\,,
\label{eq:Rg:fit}
\end{equation}
where $R_0$ and $\nu$ are the fitting parameters.
From the data in figure \ref{fig:nu} we estimate in the low temperature limit
\begin{equation}
\nu \ = \  0.348 \pm 0.007 
\label{valnu}
\end{equation}
This is so close to the mean field value $\nu = 1/3$ of the collapsed phase, that obviously  we are
in that phase.

The parameter $R_0$ that we present in Fig.~\ref{fig:R0}  describes the {\it effective} distance 
between the monomers.
In the collapsed phase we estimate  in the low temperature limit
\[
R_0 \ = \ 3.25 \pm 0.15  \ \  ({\buildrel _{\circ} \over {\mathrm{A}}})
\]
This  is clearly smaller than  the bare value (\ref{cond1}) in our model,
proposing that in the collapsed phase the monomers have the
tendency to become more densely packed also  {\it along} the polymer chain.

From our data we are not able to deduct any non-vanishing value for the sub-leading 
critical exponent $\Delta_1$
in (\ref{nu}). 

When the temperature increases beyond $\log_{10}T \approx 0$, $\nu(T)$ starts increasing
and we enter  a transition region between the collapsed phase
and the putative random walk phase. At the same time as the value of $\nu$ starts increasing,
the value of $R_0$ decreases and when temperature approaches the value \cite{foot1}
\begin{equation}
T_{PDB} \ \approx \  3.81 \pm 1.52
\label{logT}
\end{equation}
we obtain the fit
\begin{equation}
R_{g} \ \approx \ 2.8 \, \cdot  N^{0.38}
\label{fitulf}
\end{equation}
for $50 \leq N \leq 1.800$ which is very close to the estimate  \cite{ulf}
\begin{equation}
{R_g}^{\hskip -0.15cm PDB} \ \approx \ 2.25 \, \cdot N^{0.38}
\label{fitpdb}
\end{equation}
that describes the dependence of the radius of gyration on the number $N$ of $C_\alpha$ carbons for all single
strand proteins in PDB with $75 < N < 1.000$. This suggests that the model probably gives its best approximation to the
PDB data in its collapsed phase, near the transition to the random walk  phase. However, we point out that when 
 $T \approx T_{PDB}$ both $\nu(T)$ and $R_0(T)$ have a quite strong temperature dependence, indicative of vicinity
 of a phase transition that makes the accuracy of our estimates prone to
 relatively large errors, and for more precise estimates one needs simulations with substantially more computer time.

\subsection{ RW and SARW phases}

\begin{figure}[!thb]
\begin{center}
\includegraphics[scale=0.43,clip=true]{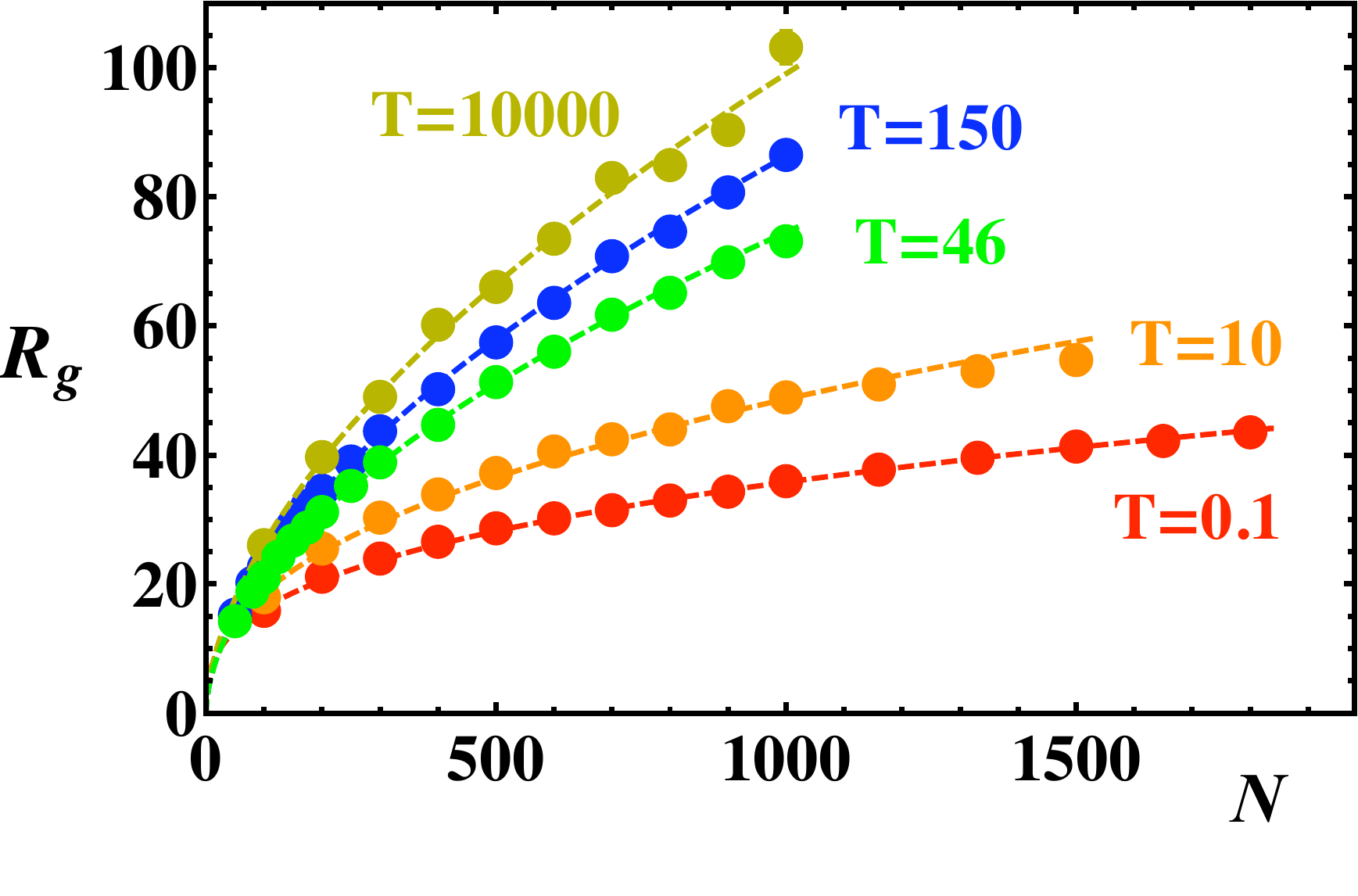}
\end{center}
\caption{The radius of gyration $R_g$ {\it vs.} the number of monomers $N$ at different values of
temperature $T$. The error bars are of the same order or smaller than the size of the symbols used.
The dashed lines represent the fits by equation~\eq{eq:Rg:fit}. }
\label{fig:Rg:fit}
\end{figure}

In Figure \ref{fig:Rg:fit}  we display  how the radius of gyration depends on the number of monomers $N$ for
a range of values of temperature beyond the collapsed phase, and compare the data with a fit of the
form (\ref{eq:Rg:fit}). As visible in this figure, even beyond the collapsed phase  the data can be fitted with 
very high accuracy by the relation (\ref{eq:Rg:fit}). However,  unlike in the collapsed phase where the radius of gyration is
practically temperature independent,
both in the putative random walk phase 
and in the putative self-avoiding random walk phase the radius of gyration is a slowly but monotonically
increasing function of the temperature.

The transition from the collapsed phase to the putative random walk phase is  very visible in our figures
\ref{fig:nu} and \ref{fig:R0}. There is a clear, rapid transition in both $\nu(T)$ and $R_0(T)$, reminiscent of
a phase transition.   From figure \ref{fig:nu} we estimate that at the transition point $\nu$ is very close to
the value
\[
\nu \approx \frac{1}{2}
\]
which is the mean field value for the $\Theta$-point.  For $T > T_\Theta$,  the compactness index  $\nu(T)$ is  a 
slowly increasing function  of temperature that eventually plateaus around the value
\[
\nu \approx 0.58
\]
This is slightly above the $\Theta$-point value, but slightly below the SARW values reported in   \cite{Zinn},   \cite{sokal}. 
Since the compactness index $\nu(N)$ 
appears to have a tendency to approach its large-$N$ limit from above \cite{sokal}, we conclude that we are in the RW phase.

For the effective monomer distance $R_0$ we find the value
\[
R_0 \ = \ 1.67 
\pm 0.03 \ \ (\dot A)
\]
which is {\it clearly} lower than the bare value (\ref{cond1}).

In general one expects that the transition between the collapsed phase and the RW phase is a phase transition,
while the transition between the RW and SARW phases is a smooth cross-over \cite{degennes1}. The results in figures  
\ref{fig:Rg:3d}-\ref{fig:R0} are in line with this, the transition between the RW phase and the 
putative self-avoiding random walk phase is much less dramatic than the transition between the collapsed phase 
and the RW phase. This also makes the precise
identification of the RW and SARW phases more involved:

We find that asymptotically  at very high temperatures $\nu$ approaches the value 
\[
\nu
{\buildrel _{T\to\infty} \over {\longrightarrow} }
%\ \buildrel{T \to \infty}{\longrightarrow} 
\ 0.62 \pm 0.03  
\]
This is slightly above the mean field value and the final values obtained in \cite{Zinn}, \cite{sokal}, but fully in line with the computations in
\cite{sokal} that revealed that the asymptotic value of $\nu$  is reached from above as the 
number of monomers increases; We note that here we have restricted ourselves to consider only values
of $N$ in the range $50 - 1.000$ that are relevant for single strand  proteins, while  \cite{sokal} considered self-avoiding
walks with up to 80.000 steps. Consequently finite scaling corrections have a much stronger effect on our
estimates.
We also point out  that as $T\to \infty$ only the self-avoiding condition
(\ref{cond2}) persists. Thus, in this limit we {\it must} be in  the universality class of SARW.

We note that for the effective monomer distance $R_0$ we find in the high temperature limit the value
\[
R_0 \ = \ 1.62 
\pm 0.08 \ \ (\dot A)
\]
that is,  essentially the same as in the RW phase.

In summary, the distinction between the collapsed phase and the RW phase appears very clear in our analysis
of the compactness index, and suggests the presence of either a first or a second order phase transition. On the
other hand,  the transition from RW phase to SARW phase is much more difficult to pinpoint, and  it appears to proceed
much more like a smooth cross-over transition than a phase transition.  These observations are fully in line with
general expectations \cite{degennes1},  and we conclude that the model \cite{ulf} does indeed correctly describe 
all the three  phases of a polymer.

\section{Elastic Energy}

\subsection{General behavior}

\begin{figure}[!thb]
\begin{center}
\begin{tabular}{c}
\includegraphics[scale=0.43,clip=true]{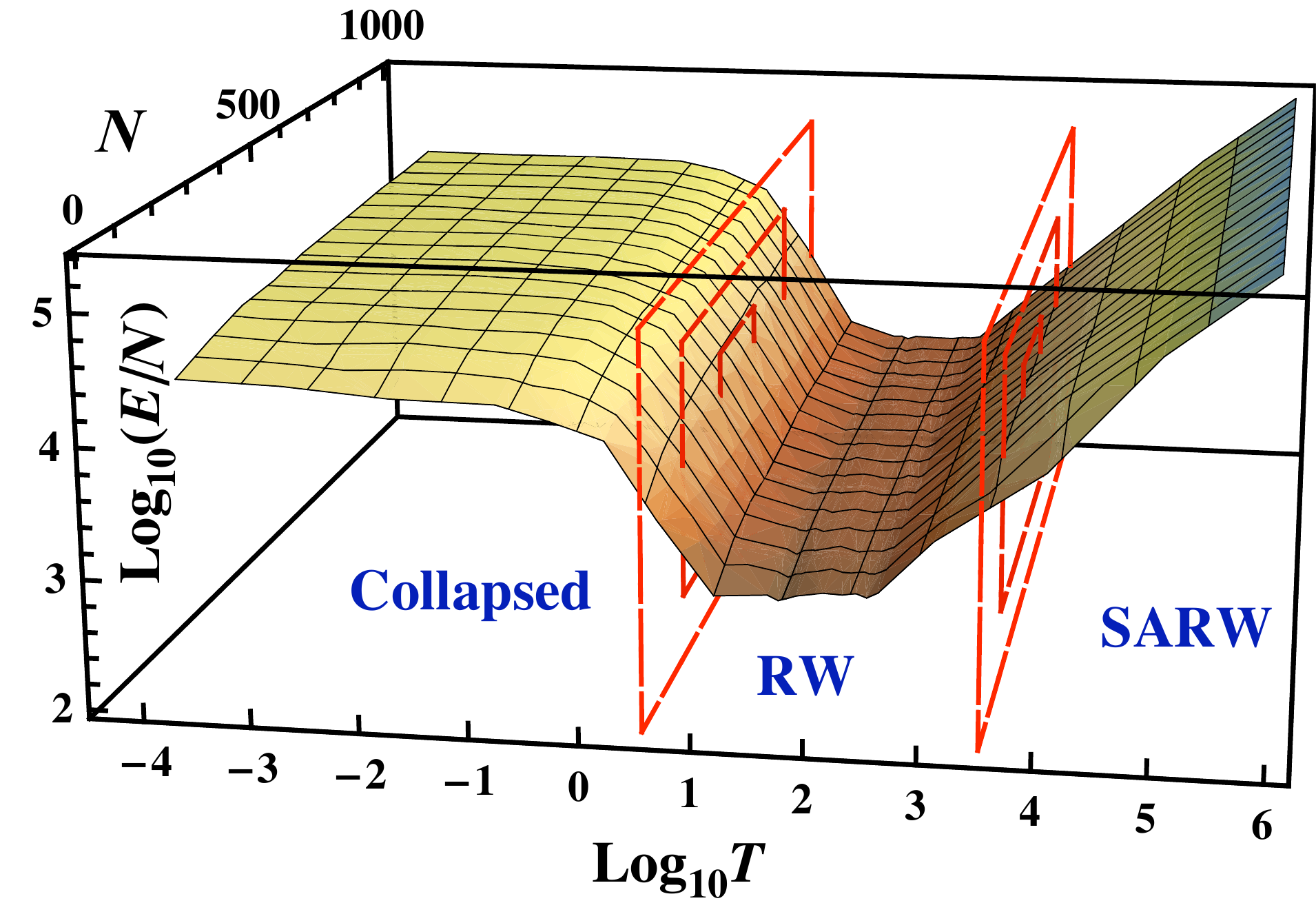}
\end{tabular}
\end{center}
\caption{A three dimensional plot of a logarithm of
the specific energy  as a function 
of a logarithm of temperature $T$ and the number of monomers $N$.
The three  phases are identified and the putative 
position of the ensuing critical temperatures are denoted by the vertical planes.}
\label{fig:energy:landscape}
\end{figure}

For the  fixed parameter values that we have given
in table \ref{tbl:parameters} the free energy (\ref{eq:model2})  is a function of two extrinsic parameters,  the temperature $T$ and 
the number of monomers $N$. Its numerical value can be identified as the {\it elastic} energy of the polymer chain.
In figure \ref{fig:energy:landscape}  we display a three dimensional plot of (a logarithm of)  the {\it specific} elastic
energy {\it i.e.} the elastic energy per monomer  as a function of these two parameters 
\begin{equation}
E_{specific} = \frac{E}{N}
\label{specific}
\end{equation}

In this figure  we clearly 
identify the presence of three different phases that are separated from each other by clearly identifiable  transition (critical) temperatures 
$T_{c1}$ and $T_{c2}$ (with $T_{c1} < T_{c2}$), and both the low temperature collapsed phase ($T< T_{c1}$) 
and the medium temperature RW phase 
($T_{c1} < T < T_{c2}$) are characterized by essentially  temperature {\it independent} specific energy.
Notice that  in the collapsed phase the specific energy has a value that is more than one order of magnitude larger than 
in the RW phase. This is understandable, as it should indeed take much more energy to extend a polymer that is collapsed and resists being
extended, than a polymer that behaves like an ideal chain and thus does not really care about its shape. 
The increase of temperature beyond $T_{c2}$  leads to a  transition to the SARW phase,
which is characterized by a power-law increase of the specific energy as  a function of the  temperature: The larger its thermal
fluctuations, the more the polymer resists to become extended.
Note also that in the collapsed and RW phases the specific energy  in figure \ref{fig:energy:landscape} 
exhibits  a weak dependence on the number of monomers $N$.  But in the high temperature SARW phase 
the specific energy  becomes essentially independent of $N$. This is consistent with
the expected behavior of  self-avoiding  random walk,   it is driven solely by the condition (\ref{cond2}) and no
reference to the details of the free energy survives  the infinite temperature limit. In this limit, the polymer is only
subject to random thermal fluctuations.

\subsection{Critical temperature}

\begin{figure}[!thb]
\begin{center}
\begin{tabular}{c}
\includegraphics[scale=0.43,clip=true]{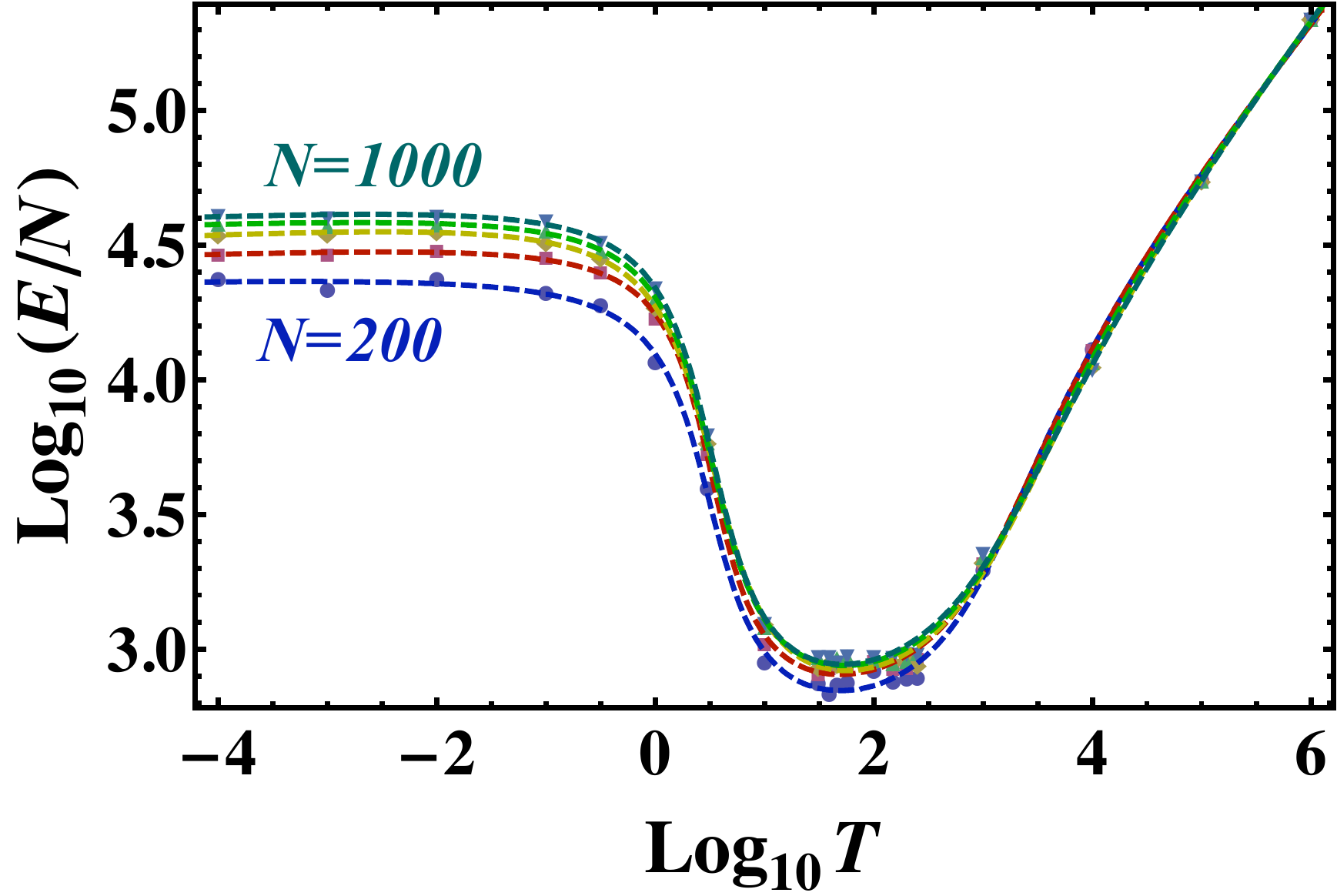}
\end{tabular}
\end{center}
\caption{The approximations (dashed lines) of the calculated numerical values (dots) 
of the specific energy by the  function \eq{eq:E:fit1}, \eq{eq:E:fit2}
when $N=200,\, 400,\, 600,\, 800,\, 1000$. The specific
energy is a monotonically rising function of the monomer number.
The lowest and the highest sets correspond to $N=200$ and $N=1000$, respectively.}
\label{fig:LogE:fit}
\end{figure}

We have found that the dependence of the specific energy 
on  temperature displayed in  figure  \ref{fig:energy:landscape} 
can be  approximated with a {\it very} good accuracy by a function
\beqn
\log_{10} (E/N) = F^{\mathrm{fit}}(\log_{10} T)
\label{eq:E:fit1}
\eeqn
that has  the following explicit form
\beqn
F^{\mathrm{fit}}(x) & = & h_1 + h_2 \arctan[h_3 (x - x_1)] \nonumber\\
& & + h_4 x \, \arctan[h_5 (x - x_2)] - h_6 x\,.
\label{eq:E:fit2}
\eeqn
The parameters $h_1 \dots h_6$ and $x_{1,2}$ are determined by fitting to the numerical
data at fixed value of the monomer number $N$. This explicit form
yields an  excellent  fit  whenever there are more than around  $N=100$ monomers.  In figure \ref{fig:LogE:fit}
we display several examples where we have fitted the functional form (\ref{eq:E:fit1}), (\ref{eq:E:fit2})
to polymers as described by our model, where  the values of $N$ are between $200$ and $1.000$. 

\begin{figure}[!thb]
\begin{center}
\begin{tabular}{c}
\includegraphics[scale=0.43,clip=true]{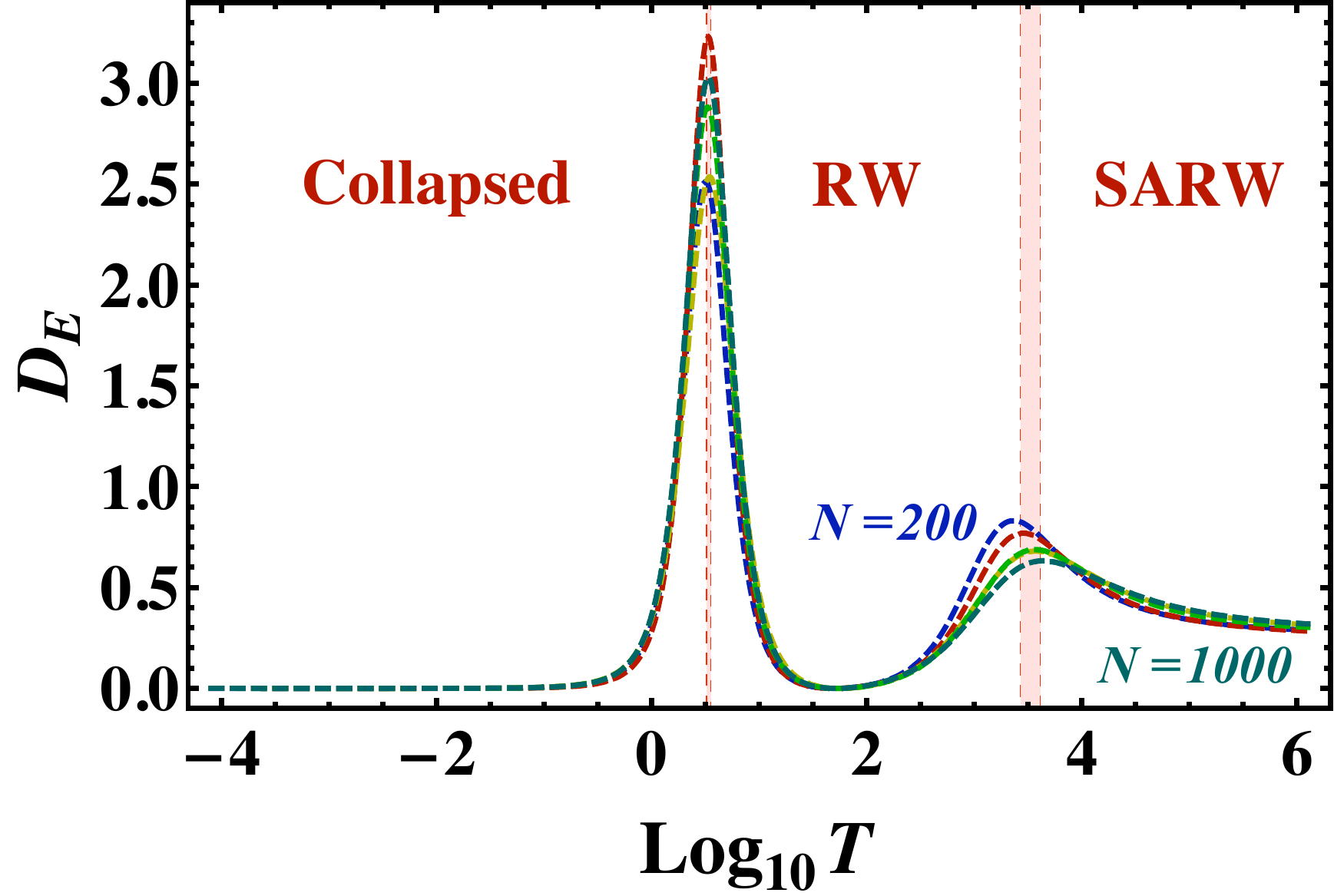}
\end{tabular}
\end{center}
\caption{The quantity~\eq{eq:DE} obtained from
the best fits of the  functions \eq{eq:E:fit1} and \eq{eq:E:fit2} for various 
values of monomer length $N$. The vertical red lines correspond to the critical temperatures
\eq{eq:Tc1} and \eq{eq:Tc2}. The width of the lines gives the uncertainty in the definition 
of the critical temperatures.}
\label{fig:finding:Tc}
\end{figure}

The  fitted functional form (\ref{eq:E:fit1}), (\ref{eq:E:fit2}) allows us to pinpoint  
the two critical temperatures $T_{c1}$ and $T_{c2}$. For this 
we locate the maxima of the squared logarithmic derivative of the specific energy with 
respect to the logarithm of temperature, 
\beqn
D_E(T,N) = \Biggl[\frac{\partial \log_{10} E(T,N)}{\partial \log_{10} T}\Biggr]^2\,.
\label{eq:DE}
\eeqn
This quantity resembles susceptibility that  is known to have its maxima at the 
location of critical temperature(s). The result is shown in figure \ref{fig:finding:Tc}.

The maxima of  \eq{eq:DE} appear as peaks that are clearly 
visible for all values of $N$ that we have studied and displayed in figure 
\ref{fig:finding:Tc}. From the results in table \ref{tbl:critical:temperatures}
we estimate that the critical temperatures have the following values, 
\beqn
&& \log_{10} T_{c1} = 0.53 \pm 0.02\,, \quad \mbox{or} \quad T_{c1} = 3.38 \pm 15
\label{eq:Tc1}\\
&& \log_{10} T_{c2} = 3.52 \pm 0.09\,, \quad \mbox{or} \quad T_{c2} = 3306 \pm 716 \qquad
\label{eq:Tc2}
\eeqn
Notice that the position of the first maximum is practically the same for all values of $N$, 
but the larger the value of $N$ the higher the height of the maximum. This 
indicates that the transition between the collapsed phase and the RW phase at $T=T_{c1}$ 
is indeed phase transition, which is either  of the second order or of the first order; Our analysis is
not sufficient to determine the order of this transition. 

On the other hand,
the transition between the RW and SARW phases at $T=T_{c2}$ is likely to be a smooth 
crossover transition since now both the position of the maximum and its height do not reflect any
similar clearly localized  profile with increasing values of  monomer number $N$. 
\begin{table}[!htb]
\begin{centering}
\begin{tabular}{|c|c|c|}
\hline
$N$ & $\log_{10} T_{c1}(L)$ & $\log_{10} T_{c2}(L)$ \\
\hline
200&   0.5023&   3.365\\
300&   0.5114&   3.397\\
400&   0.5229&   3.450\\
500&   0.5402&   3.599\\
600&   0.5379&   3.570\\
700&   0.5671&   3.552\\
800&   0.5184&   3.563\\
900&   0.5360&   3.534\\
1000&  0.5254&   3.638\\
\hline
Avr. & {\bf 0.53(2)} & {\bf 3.52(9)} \\
\hline
\end{tabular}
\end{centering}
\caption{The critical temperatures $T_{c1}$ and $T_{c2}$, 
determined for each fixed number of monomers $N$. The
averaged value is shown in the last row (in the bold face) along 
with respective errors.
\label{tbl:critical:temperatures}
}
\end{table}

\section{The phase dependence of the free energy}

We have found that in each of the three phases the elastic energy computed from (\ref{eq:model2}) has its distinct, 
universal dependence on the monomer number $N$, alternatively radius of gyration $R_g$. 

\subsection{Collapsed phase}

In the collapsed phase $T < T_{c1}$ the dependence of the free energy on the number of monomers can 
be described by the following temperature independent, logarithmically corrected linear law:
\beqn
E(N)/E_0  = C_{\mathrm{Coll}} N \ln \frac{N}{N_0^{\mathrm{Coll}}}\,.
\label{eq:E:log}
\eeqn
Here $E_0$ is a parameter that  defines the scale of the energy (say) in electronvolts and must be obtained by an independent measurement.
We find the presence of the logarithmic correction to scaling - as opposed to the 
analytic corrections proposed by (\ref{genobs}) - to be quite notable: We have made a very detailed analysis
of the functional form (\ref{eq:E:log}) and the logarithmic correction to scaling is consistently exceeding the accuracy
of any power-law alternative.

\begin{figure}[!thb]
\begin{center}
\begin{tabular}{c}
\includegraphics[scale=0.43,clip=true]{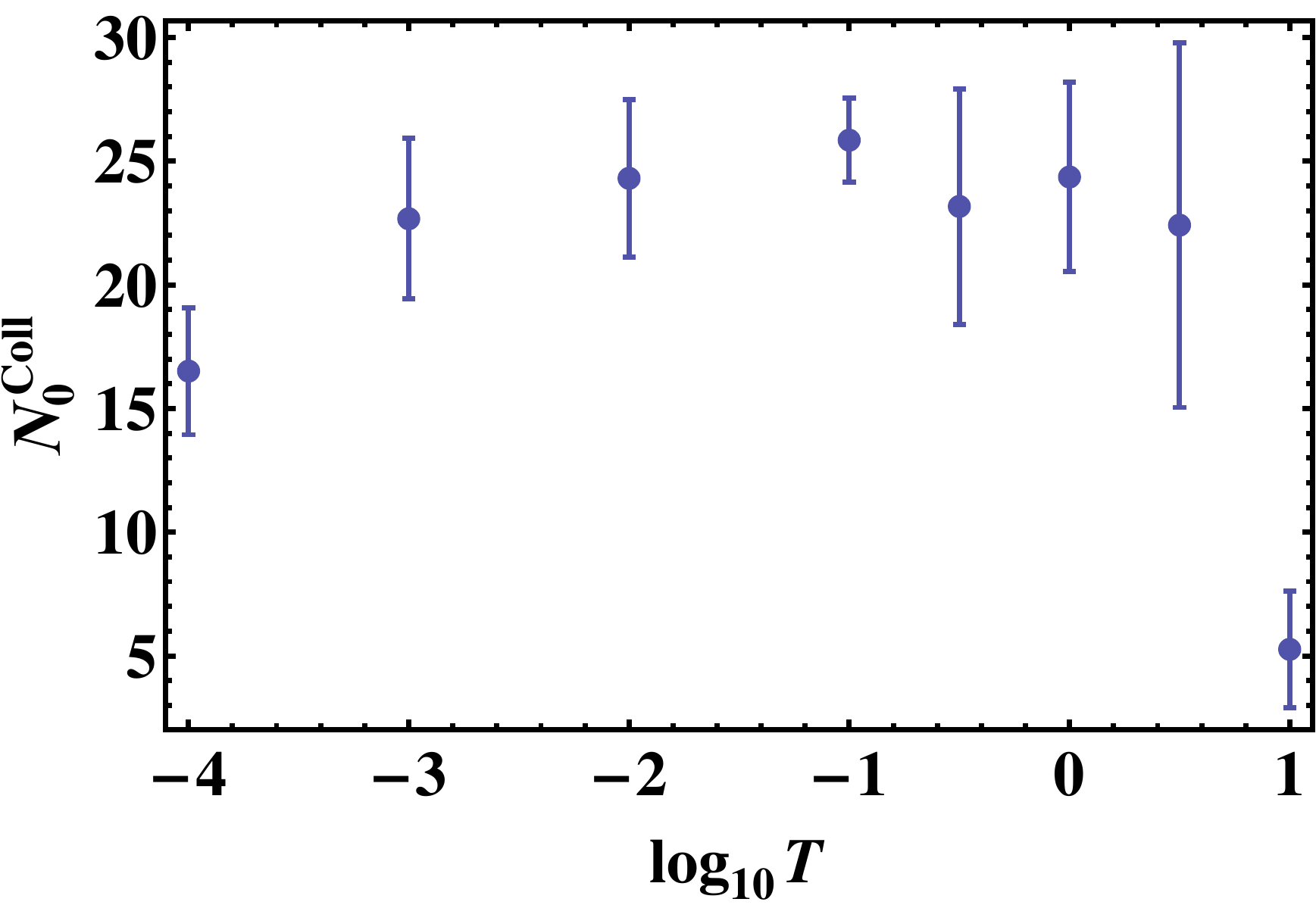}\\[3mm]
\includegraphics[scale=0.43,clip=true]{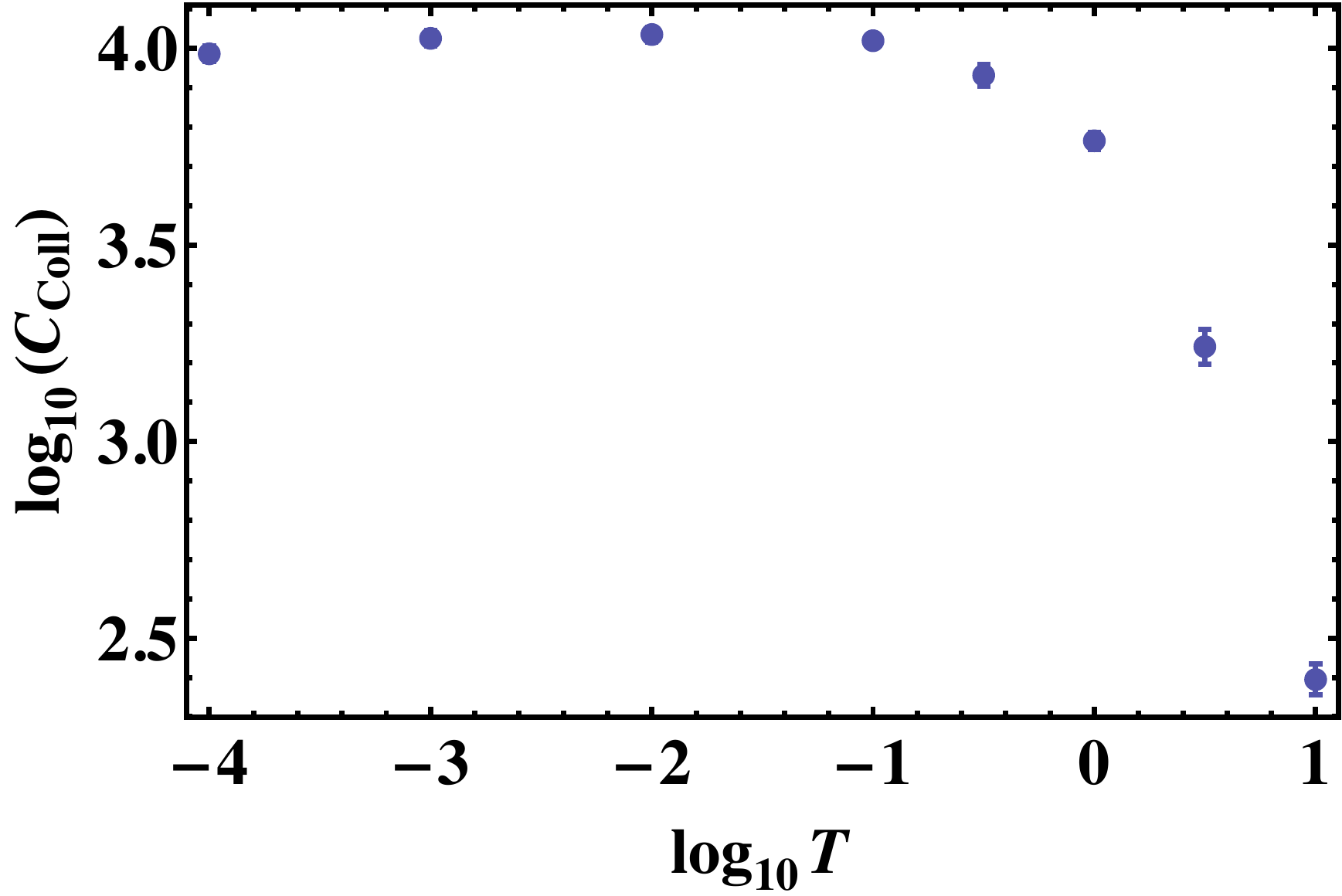}
\end{tabular}
\end{center}
\caption{The parameters of the fit~\eq{eq:E:log}: $N_0^{\mathrm{Coll}}$ (the upper plot) and $C_{\mathrm{Coll}}$ (the lower plot).}
\label{fig:L0.Coll.fit}
\end{figure}

The parameters $N_0^{\mathrm{Coll}}$ and $C_{\mathrm{Coll}}$ can be calculated using a fitting procedure. 
The results  are shown, respectively, in figure \ref{fig:L0.Coll.fit}. The  parameter $N_0^{\mathrm{Coll}}$
is essentially temperature independent in the low-temperature regime,  with value
\[
N_0^{\mathrm{Coll}} \simeq 22
\]

In terms of the radius of gyration we get from (\ref{eq:Rg:fit}), (\ref{valnu}) the approximate expression (per units of energy)
\begin{equation}
E(R_g)  \ \approx  \ 971.0 \cdot  R_g^{2.86} \cdot  \ln \left[ \frac{R_g}{9.53} \right] 
\label{ERg1}
\end{equation}
The relevant aspect of (\ref{ERg1}) is its dependence on $R_g$. Since the radius of gyration scales in proportion to the end-to-end distance the result
(\ref{ERg1}) means there is a very rapidly growing
elastic force between the end points of the collapsed polymer in our model, in particular the elastic force is growing 
clearly more rapidly than in Hooke's law.

Notice that  according to the value of the critical temperature~\eq{eq:Tc1}, 
the last data point in Fig.~\ref{fig:L0.Coll.fit} (the one with the highest temperature value) 
is  in the RW phase. However, we have found that the two parameter fit ~\eq{eq:E:log} can still be 
applied  to successfull describe this  point. 
 
\subsection{RW phase}

\begin{figure}[!thb]
\begin{center}
\begin{tabular}{c}
\includegraphics[scale=0.43,clip=true]{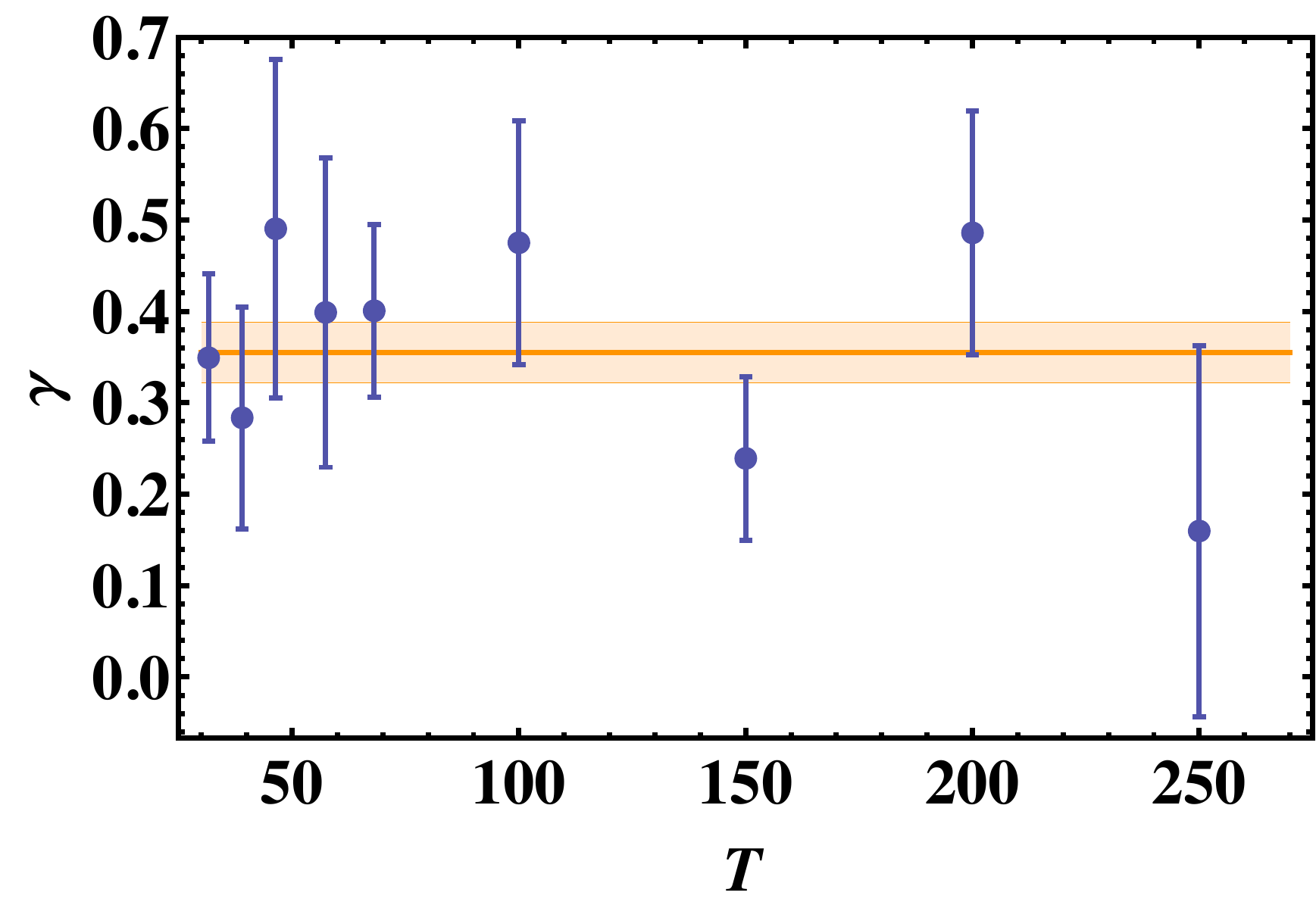}\\[3mm]
\includegraphics[scale=0.43,clip=true]{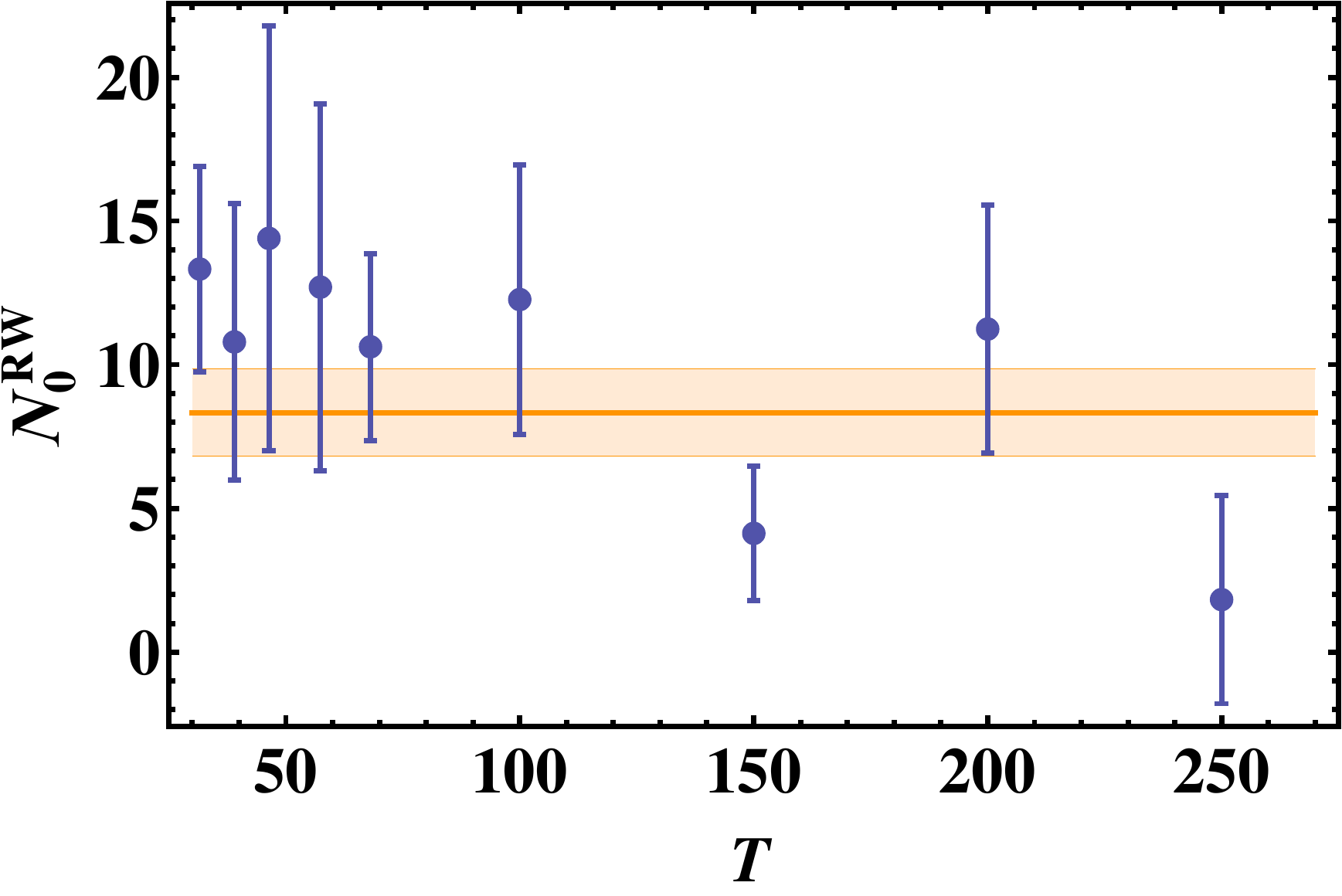}\\[3mm]
\includegraphics[scale=0.43,clip=true]{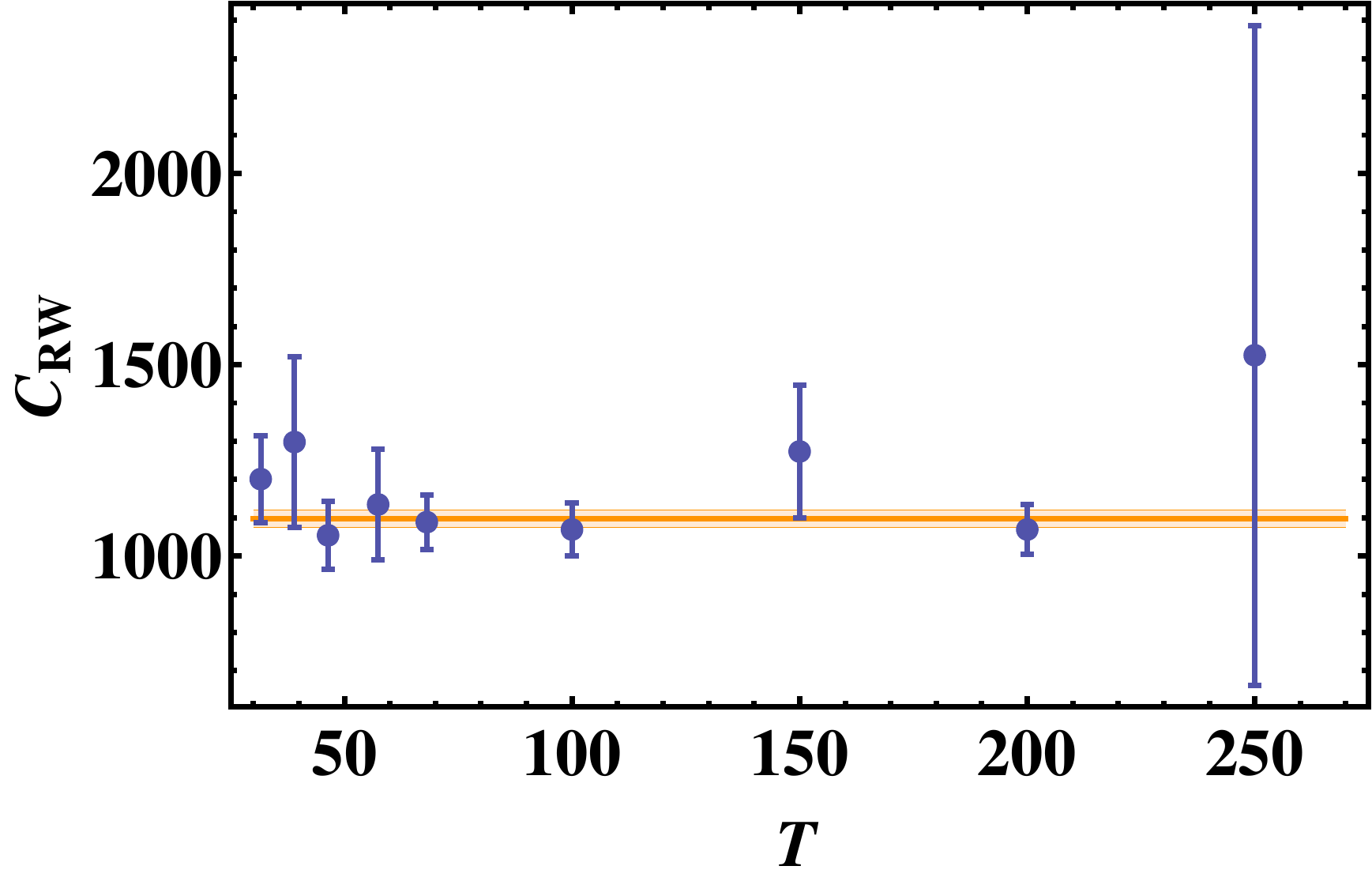}
\end{tabular}
\end{center}
\caption{The best fit parameters $\gamma$, $N_0^{\mathrm{RW}}$ and $C_{\mathrm{RW}}$ 
of the function~\eq{eq:E:power}.
The horizontal lines mark the central values~\eq{eq:RW:central},
and the width of the lines describe  the corresponding errors.}
\label{fig:RW:best:fits}
\end{figure}

In the RW phase we have found that the energy obeys the  following  scaling law (per units of energy)
\beqn
E(T,N) = C_{\mathrm{RW}}(T) N \Biggl[1 - \left(\frac{N}{N_0^{\mathrm{RW}}(T)}\right)^{-\gamma(T)}\Biggr]\,.
\label{eq:E:power}
\eeqn
This  is an example of the general form (\ref{genobs}).
The best fits of the  parameters $\gamma$, $N_0^{\mathrm{RW}}$ and $C_{\mathrm{RW}}$ are shown in 
figure~\ref{fig:RW:best:fits} as functions of temperature.  We find that all of these parameters are essentially 
temperature independent with the following average central values, 
\begin{equation}
\begin{matrix}
\gamma & = & 0.355(33) \\ 
N_0^{\mathrm{RW}} & = & 8.3(1.5) \\
C_{\mathrm{RW}} & = & 1098(22) \end{matrix}
\label{eq:RW:central}
\end{equation}
These values are shown as horizontal lines  in figure~\ref{fig:RW:best:fits}.

If we use the approximation that $\nu \approx 1/2$ in the RW phase, (\ref{eq:RW:central}) 
gives us the Hooke's law with a (temperature dependent) correction term (per units of energy),
\begin{equation}
E(R_g,T) \ \approx  \  C_{\mathrm{RW}} \left( \frac{R_g}{R_0}\right) ^{2} \Biggl[1 - (N_0^{\mathrm{RW}})^\gamma \cdot 
\ \left(\frac{R_g}{R_0}\right)^{-2 \gamma}\biggr]
\label{hooke}
\end{equation}

\subsection{SARW phase}

In the SARW phase we conclude that the energy is a linear function of the monomer number (per units of energy),
\beqn
E(N,T) = C_{\mathrm{SARW}}(T) N \,.
\label{eq:E:linear}
\eeqn
and with the mean field value of the compactness index $\nu = 3/5 $ we get in terms of radius of gyration (per units of energy)
\begin{equation}
E(R_g,T) \ \approx \ C_{\mathrm{SARW}}(T) \left( \frac{R_g}{R_0} \right)^{5/3} 
\label{lastE}
\end{equation}
From our data we are not able to observe any of the correction terms in (\ref{genobs}). 
The only fitting parameter, $C_{\mathrm{SARW}}(T)$, is shown in Fig.~\eq{fig:C:SARW} as a function of temperature.
\begin{figure}[!thb]
\begin{center}
\includegraphics[scale=0.43,clip=true]{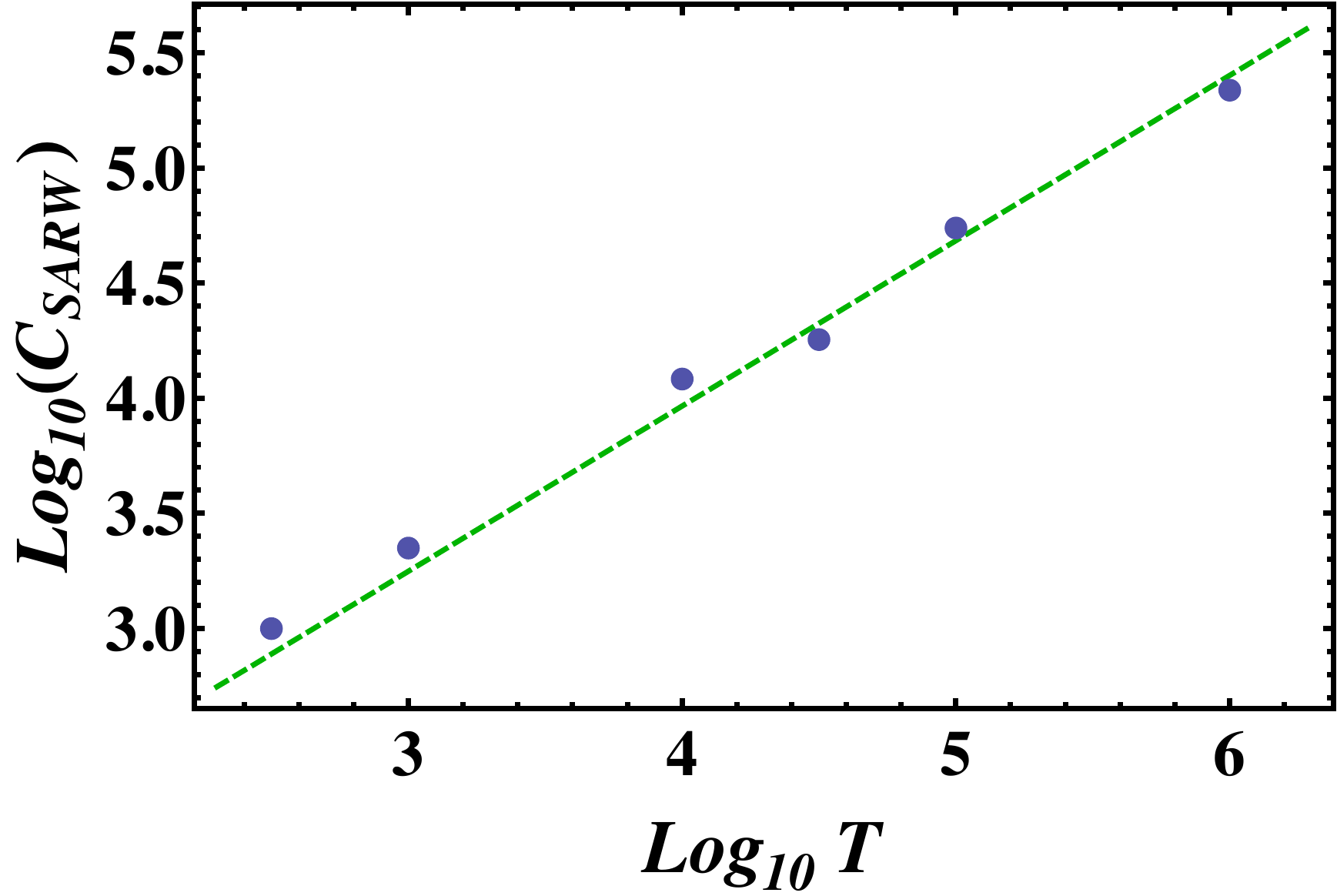}
\end{center}
\caption{The coefficient of the linear law~\eq{eq:E:linear} as  a function of temperature.
The dashed line illustrates the best fit~\eq{eq:C:SARW:fit} with the parameters~\eq{eq:C:SARW:par}.}
\label{fig:C:SARW}
\end{figure}

We also find that the temperature dependence of the coefficient $C_{\mathrm{SARW}}$ can  be described by a power law
\beqn
C_{\mathrm{SARW}}(T) = C_0\, T^\alpha\,,
\label{eq:C:SARW:fit}
\eeqn
where the prefactor $C_0$ and the exponent $\alpha$ are
\begin{equation}
\begin{matrix}
C_0 & = & 12(4)\\ 
\alpha & = & 0.72(6) \end{matrix}
\label{eq:C:SARW:par}
\end{equation}

Note that according to
the value of the critical temperature~\eq{eq:Tc2}, in Figure \ref{fig:C:SARW} the first two points that have
the lowest temperature values  belong to the RW phase but they can still be described with the present fit.
In fact, the $N$ dependence of the free energy at these two temperature values 
can be fitted {\it both} by the linear law~\eq{eq:E:linear} and by the more general power 
law~\eq{eq:E:power}. However, the power-law fit will lead to very large error bars for the best fit parameters, and therefore we have not
shown these points in Fig.~\ref{fig:RW:best:fits}. Moreover, since we expect that the transition between the RW and SARW phases is a crossover,
there should be no clear distinction between these phases in the vicinity of the transition region.

The  logarithmic~\eq{eq:E:log}, power~\eq{eq:E:power} and linear~\eq{eq:E:linear} fits  are all shown  in figure~\eq{fig:fits}.
\begin{figure}[!thb]
\begin{center}
\includegraphics[scale=0.43,clip=true]{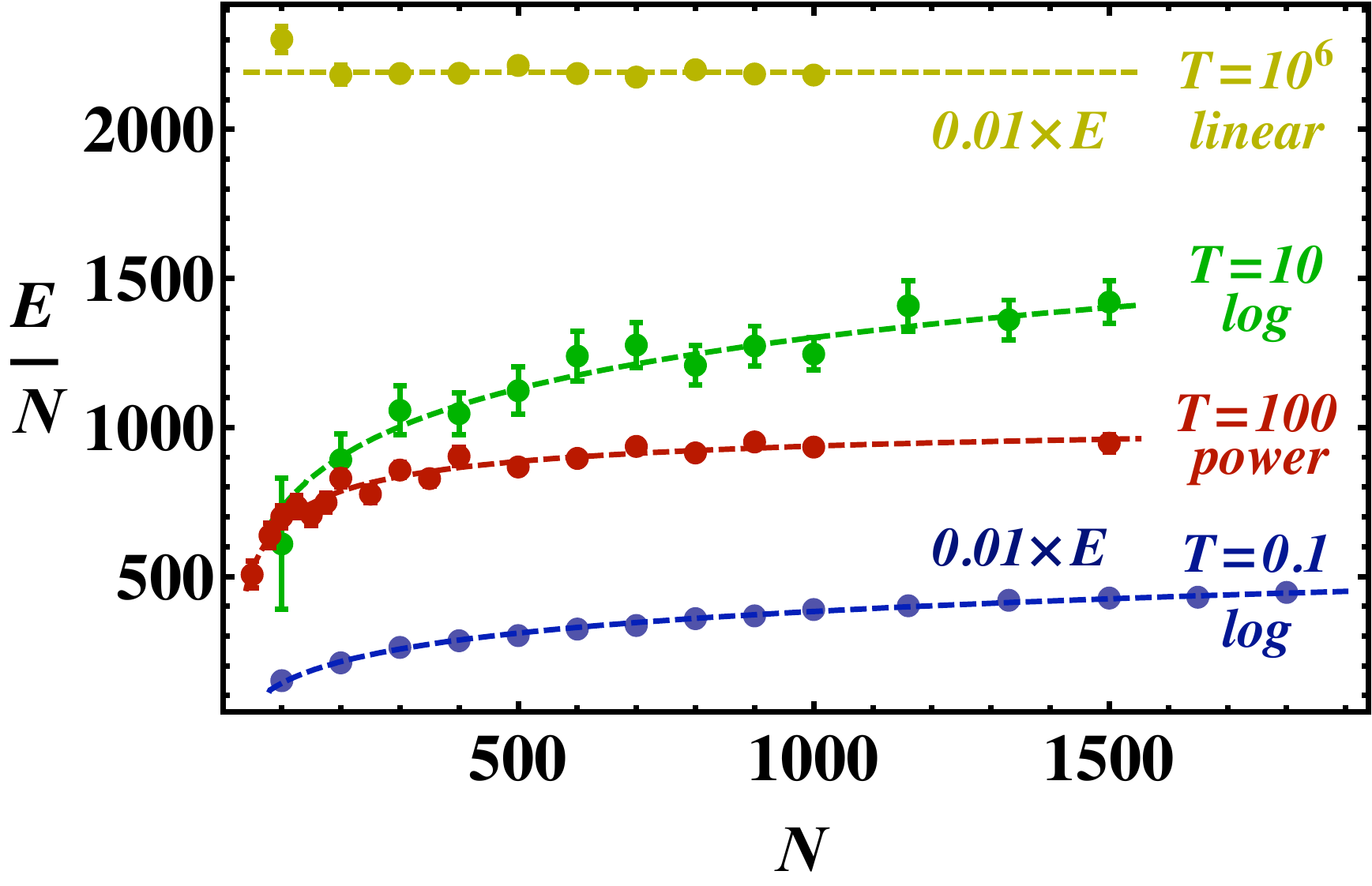}
\end{center}
\caption{The logarithmic~\eq{eq:E:log}, power~\eq{eq:E:power} and linear~\eq{eq:E:linear} at various temperatures.}
\label{fig:fits}
\end{figure}

\begin{figure}[!thb]
\begin{center}
\begin{tabular}{c}
\includegraphics[scale=0.43,clip=true]{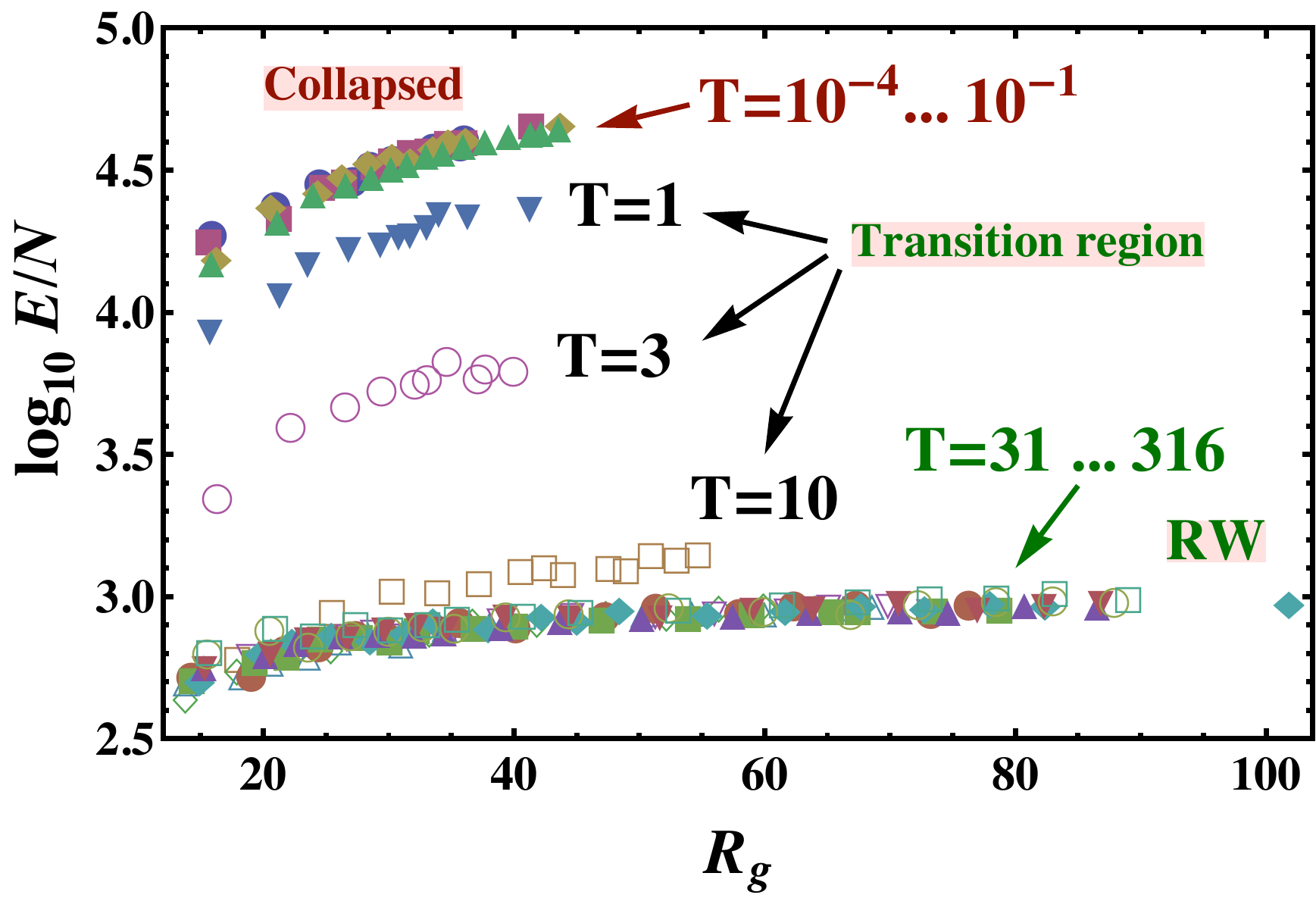} \\
\includegraphics[scale=0.43,clip=true]{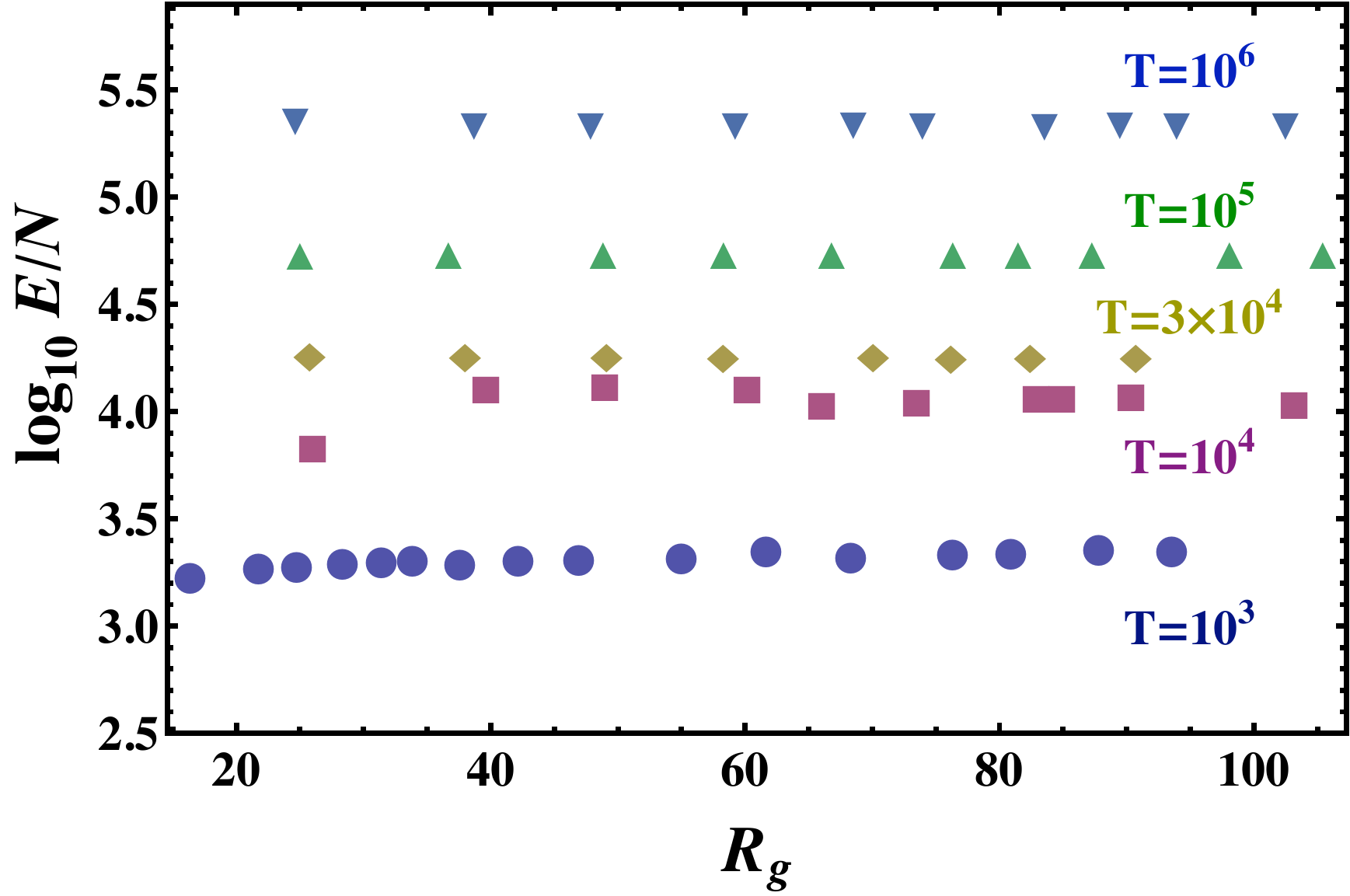} \\
\end{tabular}
\end{center}
\caption{(Logarithm of) Specific elastic energy $E/N$ {\it vs.} the radius of gyration $R_g$ at different temperatures. The
distinct points in the same series correspond to different numers of monomers $N$.
Upper plot: the low-temperature collapsed phase and medium-temperature RW phase including the
transition region between them. Lower plot: the high-temperature SARW phase. The results are fully in line with  the analytic 
expressions (\ref{ERg1}), (\ref{hooke}) and (\ref{lastE}), respectively.
}
\label{fig:univ}
\end{figure}

Finally, we summarize the results in Fig.~\ref{fig:univ} where 
we show  how the {\it specific} elastic   energy (\ref{specific})  depends on the radius of gyration $R_g$ for various temperatures.
The upper plot of Fig.~\ref{fig:univ} corresponds to the collapsed and RW phases.  It is very visible that both in the collapsed
phase and the RW phase  the relation $E_{specific} =E_{specific} (R_g)$ is indeed universal: there is no observable temperature dependence. 
We also note the rapid change from collapsed phase to RW phase.
The lower plot of Fig.~\ref{fig:univ} describes the high-temperature SARW phase.  
While an increasing function of temperature, the energy now has only 
very weak (if any) dependence on the radius of gyration.

\section{Proteins and the Huang-Lei  elastic energy }

In \cite{huang1} the authors propose that  the elastic energy of folded proteins in PDB
can be described by the following phenomenological
(Huang-Lei) formula (per units of energy)
\beqn
E_{HL}(R_g,N) = a \, N^{4/5} + b \, (N \, R_g)^{1/2} + c\,  \frac{N^2}{R_g^3},
\label{eq:Huang}
\eeqn
Here  $a$, $b$ and $c$ are fitting parameters.  By minimizing the energy, the authors \cite{huang1} compute for the
compactness index the value
\begin{equation}
\nu_{HL} = 3/7
\label{hnu}
\end{equation}
{\it A priori} this suggests \cite{huang1} that  folded proteins could be  in a universality class which
is different from the known ones (\ref{nus}). 

In this Section we shall analyze the formula (\ref{eq:Huang})
in the context of our model. We find that it gives an accurate description of data in 
our model, in particular around the transition point between the collapsed phase and the RW phase where the compactness index 
grows continuously and monotonically from around $\nu \approx 1/3$ to around $\nu \approx 1/2$ over a finite temperature interval, 
due to finite scaling effects that are characteristic to a finite length chain:
The value (\ref{hnu}) corresponds to temperature value
\[
T_{HL} \ \approx \ 12.5 \pm 1.7
\] 
in our model, which suggests that we are (slightly) above the transition temperature $T_{c1}$ between collapsed and RW phases.

\begin{figure}[!thb]
\begin{center}
\begin{tabular}{c}
\includegraphics[scale=0.43,clip=true]{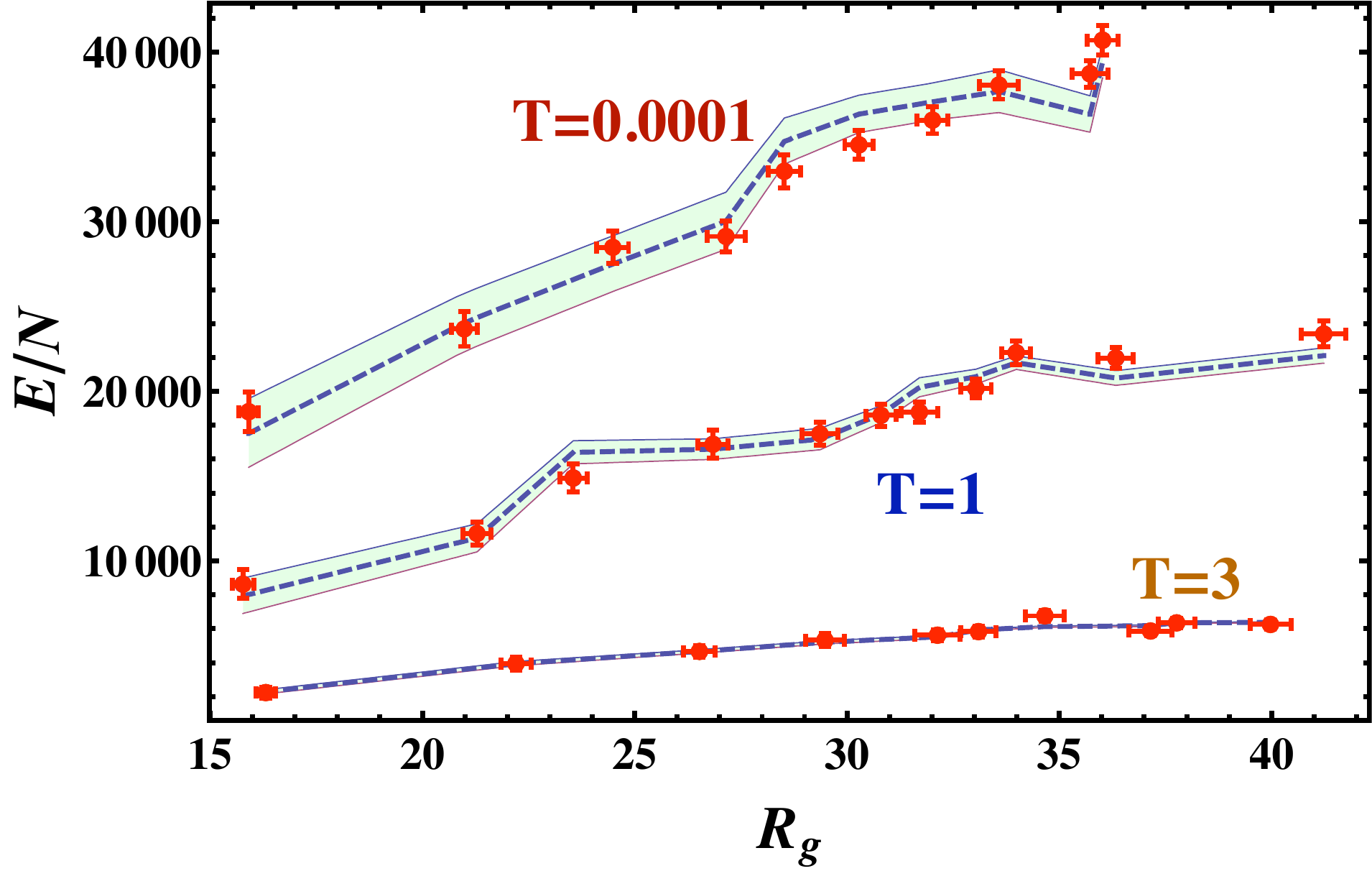}
\end{tabular}
\end{center}
\caption{ Three examples of the fits of the elastic energy $E$ by the 
Huang-Lei formula (\ref{eq:Huang}). The $T=0.0001$ line is deep in the collapsed phase while  the $T=1$ and $T=3$ lines are both in the transition region
from collapsed to RW phase, in the vicinity
of the critical value
$T_{c1}  \approx  3.38 $ .   }
\label{fig:RgE}
\end{figure}

 In figure~\ref{fig:RgE} we show examples where we have fitted (\ref{eq:Huang}) to elastic energy 
 computed from our model for three different values of the temperature: Deep in the collapsed phase and in the vicinity of the critical
 temperature $T_{c1}$ that separates the collapsed phase from the RW phase in our model. The width of the best-fit lines describes
 the uncertainty in the best-fit parameters, reflecting the statistical errors in our data. 
% \end{document}

We have found that deep in the collapsed phase the fit is not very good and consequently (\ref{eq:Huang}) does not describe fully collapsed
proteins, as expected from the value of the compactness index. 
But when we enter the transition region between collapsed phase and 
RW phase and the compactness index starts increasing (continuously as a function of temperature 
for finite length chains), 
the 
quality of the fit becomes increasingly improved and in the vicinity of the critical temperature $T_{c1}$ we find for the statistical
$\chi$-square parameter per degree of freedom $(dof)$ a value around
\[
\chi^2/(dof) \approx 1
\]
 
\begin{figure}[!thb]
\begin{center}
\begin{tabular}{c}
\includegraphics[scale=0.43,clip=true]{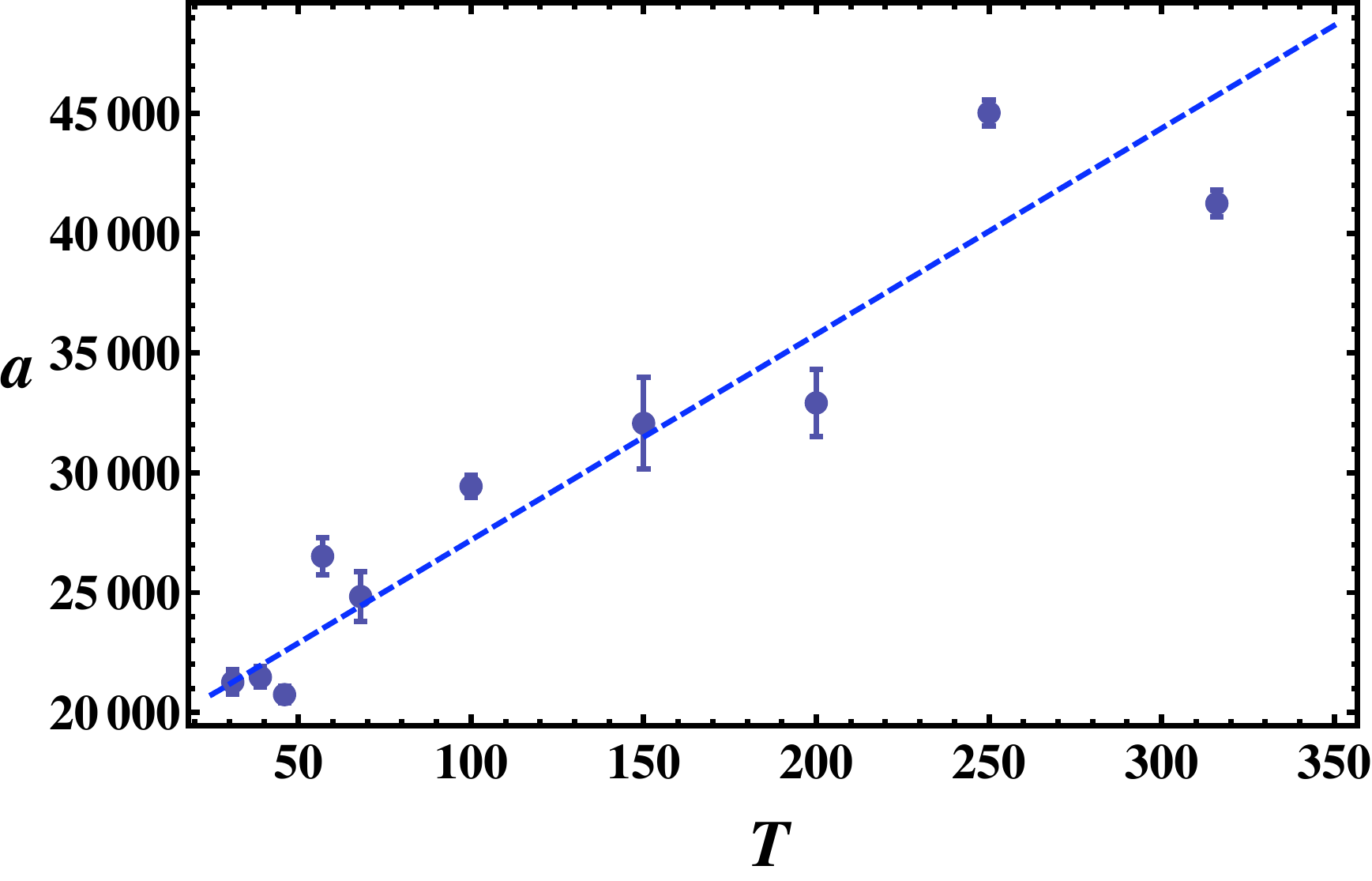} \\[3mm]
\includegraphics[scale=0.43,clip=true]{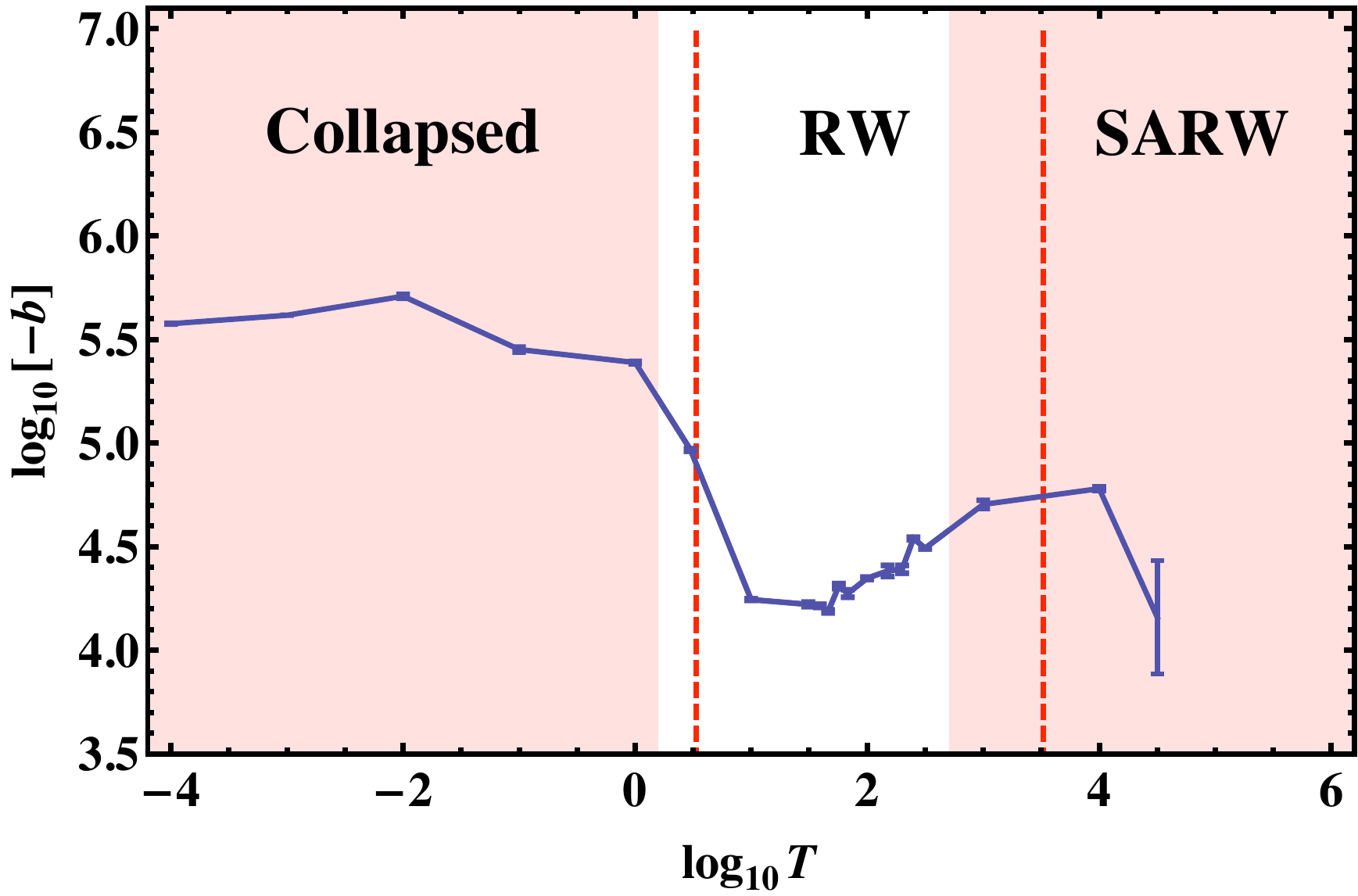} \\[3mm]
\includegraphics[scale=0.43,clip=true]{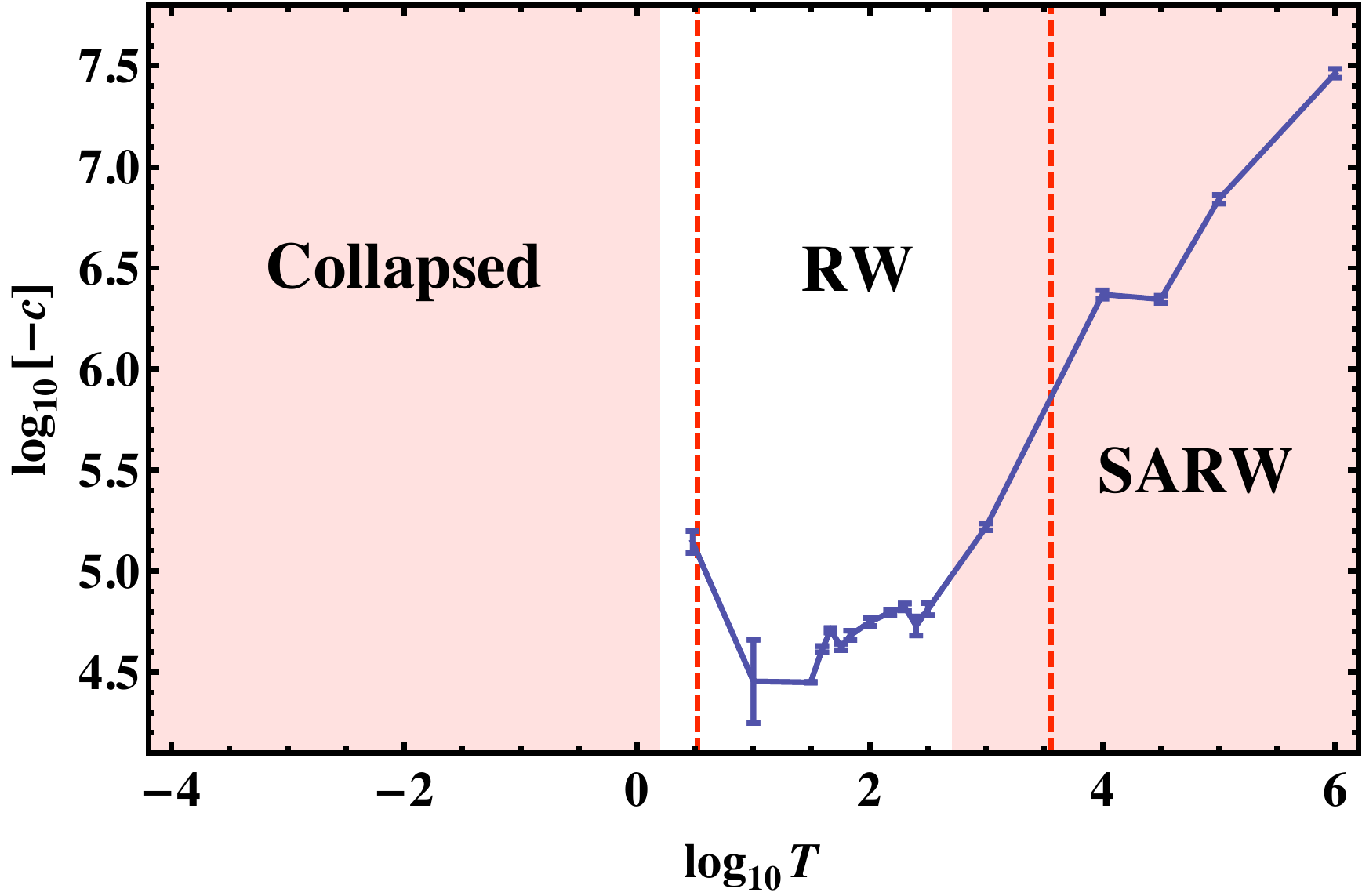} \\
\end{tabular}
\end{center}
\caption{The best fit parameters of the fits of the elastic energy~\eq{eq:Huang} are shown the log-log scale.
The description is given in the text.}
\label{fig:huang:corrected:1}
\end{figure}

In figure~\ref{fig:huang:corrected:1} we summarize our findings for  the set of best fit parameters for  \eq{eq:Huang}.
The red-colored  zones correspond to those  values of temperature where  the $\chi^2/(dof)$ parameter
is very large, typically taking values around 10 and higher. In the un-colored (white) zones 
the  $\chi^2/(dof)$ parameter has values that are in the vicinity of unity.
\begin{figure}[!thb]
\begin{center}
\begin{tabular}{c}
\includegraphics[scale=0.43,clip=true]{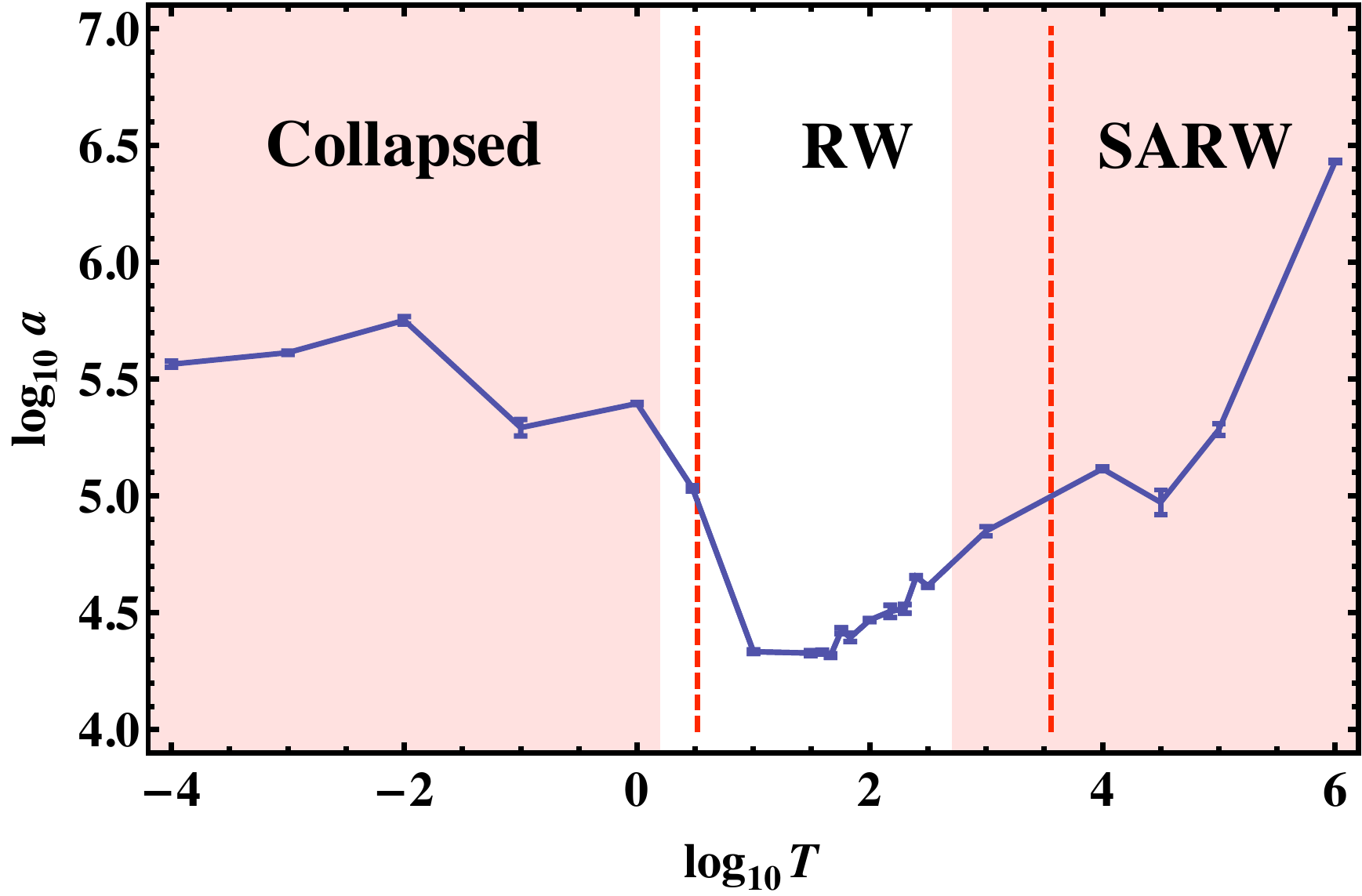} \\[3mm]
\includegraphics[scale=0.43,clip=true]{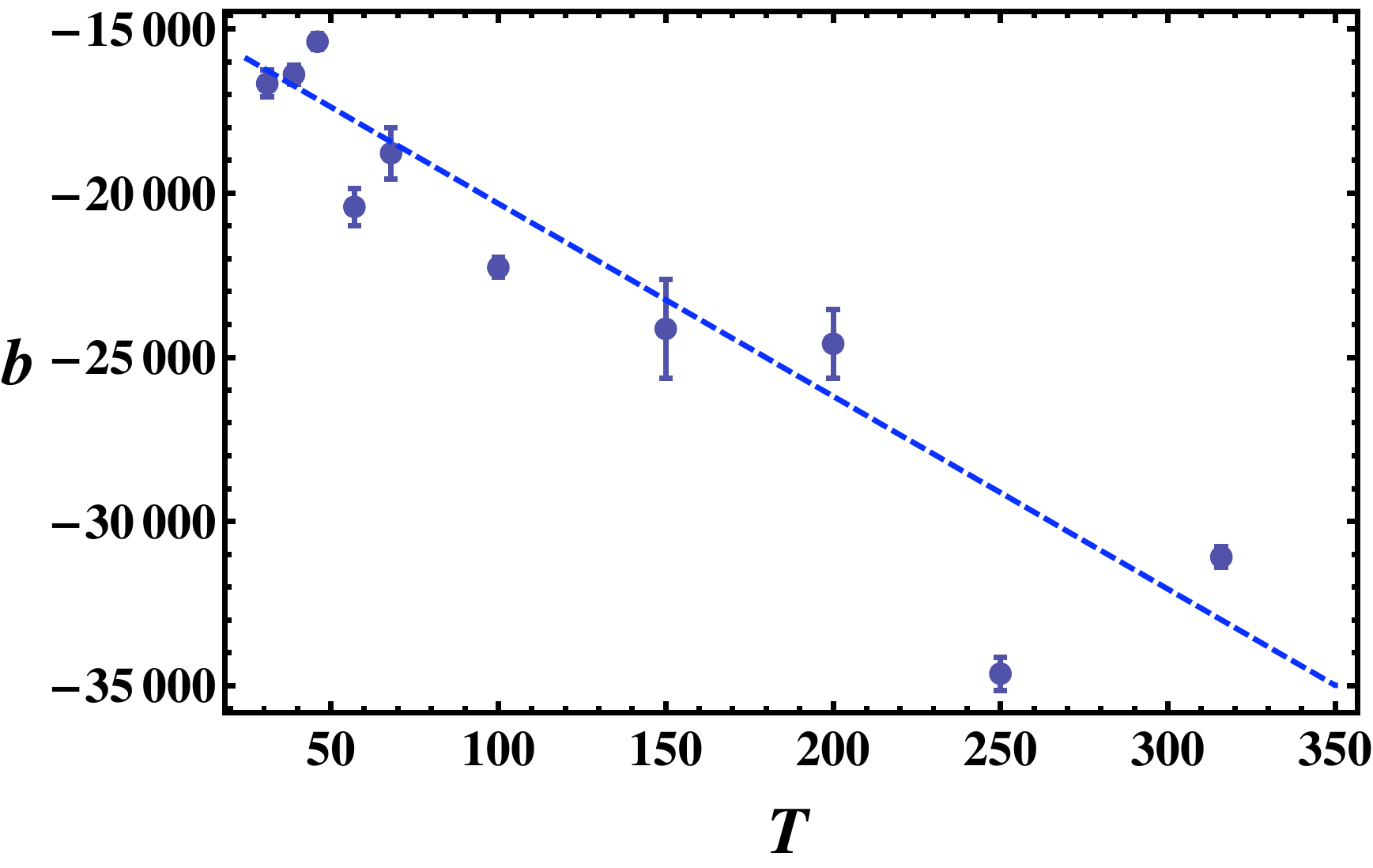} \\[3mm]
\includegraphics[scale=0.43,clip=true]{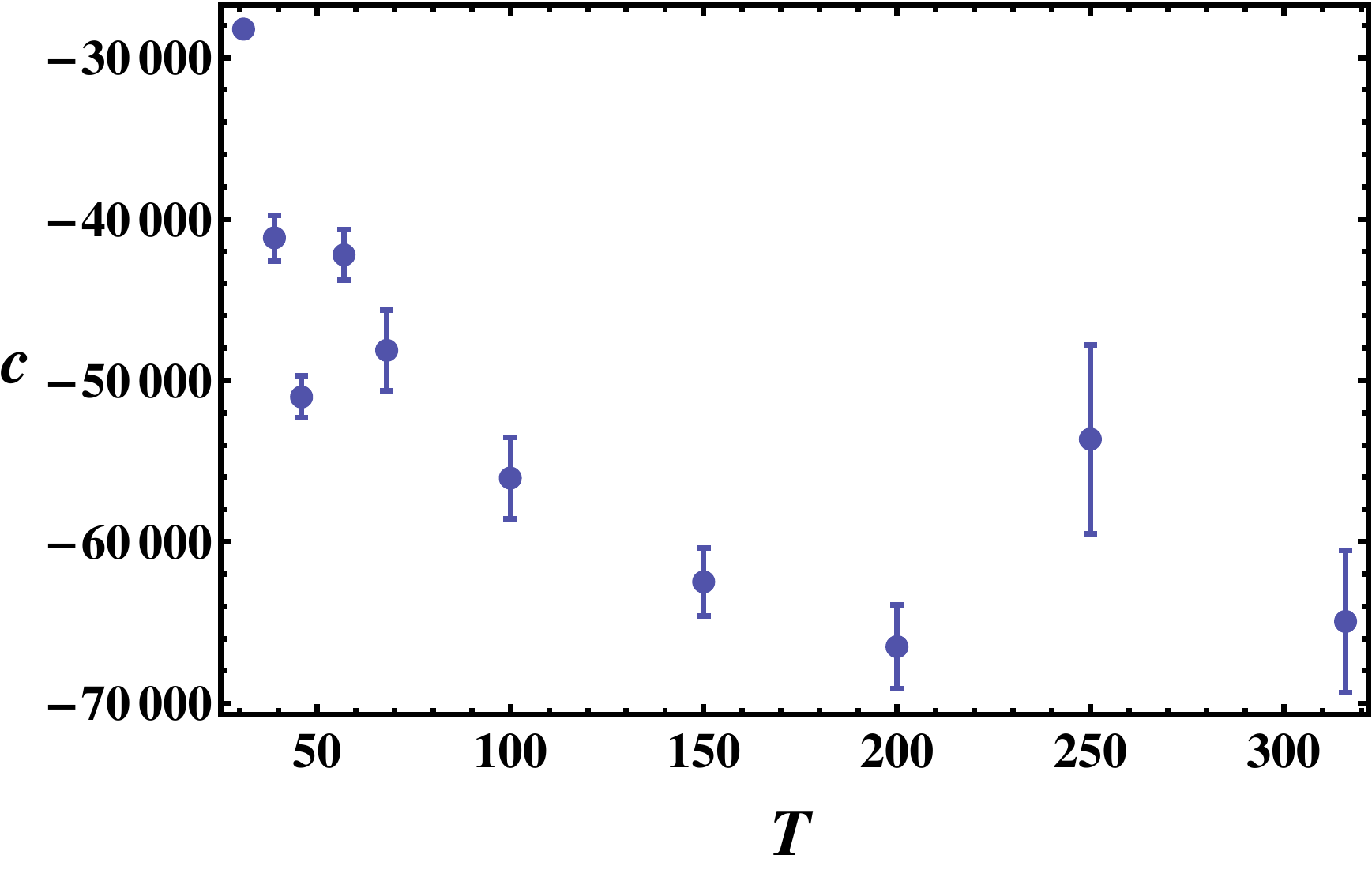} \\
\end{tabular}
\end{center}
\caption{The parameters of the fit~\eq{eq:Huang} in the RW region.
The dashed lines give are the best fits given by Eqs.~\eq{eq:ab}, \eq{eq:ab:par1}, and \eq{eq:ab:par2}.}
\label{fig:huang:local}
\end{figure}

In figure~\ref{fig:huang:local} we show the behavior of the parameters $a$, $b$ and $c$ 
in the region where the  $\chi^2/(dof)$ values are in the vicinity of unity that is near the transition between collapsed phase  and RW phase,
and within  the RW phase. We have found  that the temperature dependence of the 
parameters $a$ and $b$ can be fitted by linear functions:

\beqn
a(T) = C_a + \Biggl(1+\frac{T}{T_a}\Biggr)\
\\
b(T) = C_b + \Biggl(1+\frac{T}{T_b}\Biggr)
\label{eq:ab}
\eeqn
where
\beqn
& & C_a = 8.6(1.4)\cdot 10^3\,, 
\quad \ \ \ 
T_c = 216(37)\,,
\label{eq:ab:par1}\\
\qquad
& & C_b = -1.4(1) \cdot 10^4\,,
\qquad
T_b = 246(44)\,.
\label{eq:ab:par2}
\eeqn
These fits are shown in Fig.~\ref{fig:huang:local} by the dashed lines.

Our conclusion is that the Huang-Lei formula 
\eq{eq:Huang} gives a very good description of the elastic energy in our model, in particular  when 
we are very near the transition point between the collapsed and RW phases, 
and slightly  inside the RW phase. But  it is not very accurate for temperature values 
that are deep in the collapsed phase, nor when we approach the cross-over to
the SARW phase. 
We note that this is consistent
with the behavior of the compactness index in our model as displayed in Fig.~\ref{fig:nu}.
When we compare the computed value (\ref{hnu}) with  Fig.~\ref{fig:nu} 
we find that this value corresponds to the transition region .

Together with
 \cite{huang1}, and the comparison between (\ref{fitulf}) and (\ref{fitpdb}), and (\ref{logT}) and (\ref{eq:Tc1}),  
 these results suggest  that our model  should describe the statistical properties of folded proteins in PDB, for temperature 
 values that are very close to the critical value $T_{c1} \approx 3.38$.

\section{discussion}

We have investigated the statistical properties of a homopolymer model that has been introduced to describe the properties of
collapsed proteins in Protein Data Bank. We have found that as a function of temperature 
the model does indeed realize the three known  phases of polymers: the collapsed phase, the random walk phase (RW), and the
self-avoiding random walk phase (SARW). Furthermore, we have found that the model predicts that the 
transition between the collapsed phase and the random walk phase is a phase transition, while the random walk and self-avoiding 
random walk phases are separated from each other by a smooth cross-over transition. These findings are in line with 
general arguments on the phase structure of polymers \cite{degennes1}. 

We have also computed the elastic energy as a function of radius of gyration {\it i.e.} end-to-end distance of a polymer. 
In the collapsed phase we have found that the energy grows faster than in Hooke's law, in the RW phase we find Hooke's law 
with temperature dependent corrections, and finally in the SARW phase we find that the dependency of energy 
on the radius of gyrations is weaker than in Hooke's law. It would be interesting to test our predictions experimentally in the
case of proteins, for example using atomic force microscopy.

Finally, we have compared our model with  a phenomenological expression that has been introduced by Huang and Lei 
to describe the elastic energy of collapsed proteins. We have found that the Huang-Lei formula gives a good effective
description of our model, in particular  when we are in the vicinity of the transition region that separates the collapsed phase from
the random walk phase. This is also consistent with our evaluation of the temperature dependence of the compactness index. 
When compared with the PDB data this suggests that statistical properties of collapsed proteins are indeed described by the present 
model in the vicinity of this transition point.

\begin{acknowledgments}
This work was supported by a STINT Institutional grant IG2004-2 025.
\end{acknowledgments}


\begin{thebibliography}{99}
\bibitem{foot1}  The error in the estimate is  large because around this point the temperature changes rapidly, 
indicating that we are near a phase transition.
\end{thebibliography}
\end{document}